\documentclass[structabstract]{aa}  
\usepackage{graphicx}
\usepackage{txfonts}
\usepackage{natbib}
\bibpunct{(}{)}{;}{a}{}{,}

\begin{document}
   \title{Multiwavelength campaign on Mrk 509}
   \subtitle{II. Analysis of high-quality Reflection Grating Spectrometer
   spectra}
\author{J.S. Kaastra\inst{1,2}
  \and C.P. de Vries\inst{1}
  \and K.C. Steenbrugge\inst{3,4}
  \and R.G. Detmers\inst{1,2}
  \and J. Ebrero\inst{1}
  \and E. Behar\inst{5}
  \and S. Bianchi\inst{6}
  \and E. Costantini\inst{1}
  \and G.A. Kriss\inst{7,8}
  \and M. Mehdipour\inst{9}
  \and S. Paltani\inst{10}
  \and P.-O. Petrucci\inst{11}
  \and C. Pinto\inst{1}
  \and G. Ponti\inst{12}
  }
  
\institute{SRON Netherlands Institute for Space Research, Sorbonnelaan 2,
           3584 CA Utrecht, the Netherlands 
	   \and
	   Sterrenkundig Instituut, Universiteit Utrecht, 
	   P.O. Box 80000, 3508 TA Utrecht, the Netherlands
           \and
           Instituto de Astronom\'ia, Universidad Cat\'olica del Norte, 
	   Avenida Angamos 0610, Casilla 1280, Antofagasta, Chile
	   \and
	   Department of Physics, University of Oxford, Keble Road, 
	   Oxford OX1 3RH, UK	
	   \and
	   Department of Physics, Technion-Israel Institute of Technology, 
	   Haifa 32000, Israel 
	   \and 
	   Dipartimento di Fisica, Universit\`a degli Studi Roma Tre, 
	   via della Vasca Navale 84, 00146 Roma, Italy 
	   \and
	   Space Telescope Science Institute, 3700 San Martin Drive, 
	   Baltimore, MD 21218, USA
	   \and
	   Department of Physics and Astronomy, The Johns Hopkins University,
	   Baltimore, MD 21218, USA
	   \and
	   Mullard Space Science Laboratory, University College London, 
	   Holmbury St. Mary, Dorking, Surrey, RH5 6NT, UK
	   \and
	   ISDC Data Centre for Astrophysics, Astronomical Observatory of the
	   University of Geneva, 16, ch. d'Ecogia, 1290 Versoix, Switzerland 
	   \and
	   UJF-Grenoble 1 / CNRS-INSU, Institut de Plan\'etologie et d'Astrophysique
	   de Grenoble (IPAG) UMR 5274, Grenoble, F-38041, France
	   \and
	   School of Physics and Astronomy, University of Southampton, 
	   Highfield, Southampton SO17 1BJ
}
\date{\today}

\abstract
{We study the bright Seyfert 1 galaxy Mrk~509 with the Reflection Grating
Spectrometers (RGS) of XMM-Newton using the RGS multi-pointing mode of
XMM-Newton for the   first time in order to constrain the properties of the
outflow in this object.}
{We want to obtain the most accurate spectral properties
from the 600~ks spectrum of Mrk~509 which has excellent statistical quality.}
{We derive an accurate relative calibration for the effective area of
the RGS, derive an accurate absolute wavelength calibration, improve the
method for adding time-dependent spectra and enhance the efficiency of the
spectral fitting by two orders of magnitude.}
{We show the major improvement of the spectral data quality due to the use of
the new RGS multi-pointing mode of XMM-Newton. We illustrate the gain in
accuracy by showing that with the improved wavelength calibration the two
velocity troughs observed in UV spectra are resolved.}
{}

 \keywords{Galaxies: active --  quasars: absorption lines -- X-rays: general
               }
\maketitle

\section{Introduction}

Outflows from Active Galactic Nuclei (AGN) play an important role in the
evolution of the super-massive black holes (SMBH) at the centres of the AGN, as
well as on the evolution of the host galaxies and their surroundings. In order
to better understand the role of photo-ionised outflows, the geometry must be
determined. In particular, our goal is to determine the distance of the
photo-ionised gas to the SMBH, which currently has large uncertainties. For this
reason we have started a large monitoring campaign on one of the brightest AGN
with an outflow, the Seyfert 1 galaxy \object{Mrk~509} \citep{kaastra2011}. The main goal of this
campaign is to track the response of the photo-ionised gas to the temporal
variations of the ionising X-ray and UV continuum. The response time immediately
yields the recombination time scale and hence the density of the gas. Combining
this with the ionisation parameter of the gas, the distance of the outflow to
the central SMBH can be determined.

The first step in this process is to accurately determine the physical state of
the outflow: what is the distribution of gas as a function of ionisation
parameter, how many velocity components are present, how large is the turbulent
line broadening, etc. Our campaign on Mrk~509 is centred around ten observations
with XMM-Newton spanning seven weeks in the Fall of 2009. The properties of the
outflow are derived from the high-resolution X-ray spectra taken with the
Reflection Grating Spectrometers \citep[RGS,][]{denherder2001} of XMM-Newton.
The time-averaged RGS spectrum is one of the best spectra ever taken by this
instrument, and the statistical quality of this spectrum  can be used to improve
the current accuracy of the calibration and analysis tools. This gives rise to
challenges for the analysis of the data. The methods developed here also apply
to other time-variable sources. Therefore we describe them in some detail in
this paper.

For this work we have derived a list of the strongest absorption lines in
the X-ray spectrum of Mrk~509, and we perform some simple line diagnostics
on a few of the most prominent features. The full time-averaged spectrum will be
presented elsewhere \citep{detmers2011}.

\section{Data analysis}

\begin{table}[!ht]
\caption[]{Observation log.}
\smallskip
\label{tab:obslog}
\begin{tabular}{ccrc}
\hline
Obs. & ID & Start date & net exposure \\
nr.  &    & & (ks) \\
\hline
 1 & 0601390201 & 15 Oct 2009 & 60 \\
 2 & 0601390301 & 19 Oct 2009 & 53 \\
 3 & 0601390401 & 23 Oct 2009 & 61 \\
 4 & 0601390501 & 29 Oct 2009 & 60 \\
 5 & 0601390601 &  2 Nov 2009 & 63 \\
 6 & 0601390701 &  6 Nov 2009 & 63 \\
 7 & 0601390801 & 10 Nov 2009 & 61 \\
 8 & 0601390901 & 14 Nov 2009 & 60 \\
 9 & 0601391001 & 18 Nov 2009 & 65 \\
10 & 0601391101 & 20 Nov 2009 & 63 \\
\hline
\end{tabular}
\end{table}

Table~\ref{tab:obslog} gives some details on our observations. We quote here
only exposure times for RGS. The total net exposure time is 608 ks. No filtering
due to enhanced background radiation was needed for these observations, as the
background was very low and stable for the full duration of our campaign.

The campaign consisted of ten different observations. Each observation was
pointed a bit differently to obtain slightly different positions of the spectra
on the detectors (the Multi-Pointing Mode of RGS; steps of 0, $\pm
15^{\prime\prime}$ and $\pm 30^{\prime\prime}$). In this way holes in the
detector, due to bad CCD columns and pixels and CCD gaps, changed position over
the spectra, allowing all spectral bins to be  sampled at some time. In addition
this procedure will limit outliers in individual wavelength bins, due to
isolated noisy CCD pixels, since the noise of these pixels will be spread over
more wavelength bins.  

After all exposures were separately processed using the standard RGS pipeline of
the XMM-Newton data analysis system SAS version 9, noise due to remaining noisy
CCD columns and pixels was decreased further using the following process. 
Two-dimensional plots of the spectral image (cross-dispersion position versus
dispersion angle) and the CCD-pulse height versus spectral dispersion were
plotted in the detector reference frame for each separate CCD node, but with all
exposures combined. Since spectral features will be smeared out in these plots,
possible CCD defects like hot or dead columns and hot or dead pixels are clearly
visible. In this way a number of additional bad columns and pixels were manually
identified for each detector CCD node and added to the SAS bad pixel calibration
file. After this, all exposures were reprocessed, using the new bad pixel
tables. 

\section{Combination of spectra}

To obtain an accurate relative effective area and absolute wavelength
calibration, we need the signal-to-noise of the combined 600~ks spectrum.
Therefore we first describe how to accurately combine the individual spectra,
and discuss the effective area and wavelength scale later.

\subsection{Combination of spectra from the same RGS and spectral order}

Detectors like the RGS yield spectra with lacking counts in parts of the
spectrum due to the presence of hot pixels, bad columns or gaps between CCDs.
This gives rise to problems when spectra taken at different times have to be
combined into a single spectrum, in particular when the source varies in
intensity and the lacking counts fall on different wavelengths for each
individual spectrum. For RGS this is for example the case when the
multi-pointing mode is used like in our present Mrk~509 campaign, or when data
of different epochs with slightly different pointings are combined, or if bad
columns have (dis)appeared between different observations due to the transient
nature of some bad pixels. 

\begin{figure}[!htbp]
\resizebox{\hsize}{!}{\includegraphics[angle=-90]{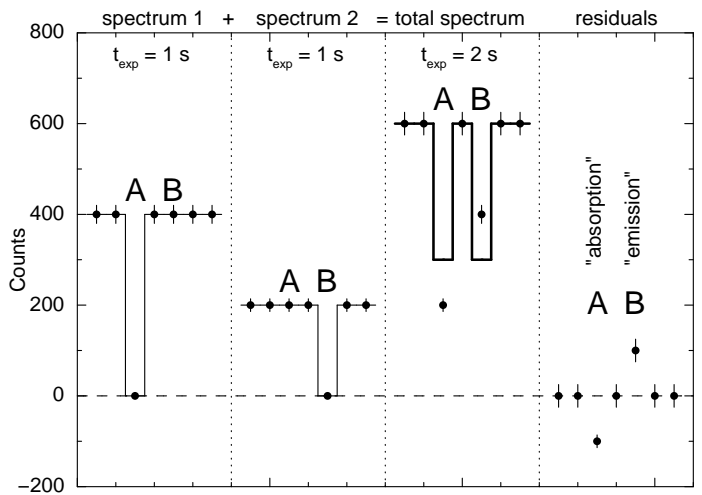}}
\caption{Illustration that spectra with different flux and different missing
bins cannot be simply added. Both spectrum 1 and 2 have 1~s exposure but have
different constant expected count rates of 400 and 200 counts\,s$^{-1}$. In
spectrum 1 pixel A is missing, in spectrum 2 pixel B. The thick solid line in
the total spectrum indicates the predicted number of counts by straightforwardly
adding the response matrices for both spectra with equal weights.}
\label{fig:pixel}
\end{figure}

Under the above conditions the SAS task {\sl rgscombine} should not be used. We
illustrate this with a simplified example (Fig.~\ref{fig:pixel}). Consider two
spectra of the same X-ray source, labelled 1 and 2,  with 7 spectral bins each
(Fig.~\ref{fig:pixel}). The source has a flat spectrum and the flux per bin in
spectrum 1 is twice the flux of spectrum 2. In spectrum 1, bin A is missing
(zero counts), and in spectrum 2, bin B is missing. The combined spectrum has
600 counts, except for bins A and B with 200 and 400 counts, respectively.
However, when the response matrices of both observations are combined, it is
assumed that bins A and B were on for 50\% of the time, hence the model predicts
300 counts for those bins. Clearly, the data show a deficit or excess at those
bins, which an observer could easily misinterpret as an additional astrophysical
absorption or emission feature at bins A and B.

How can this problem be solved? One possibility is to fit all spectra
simultaneously. However, when there are many different spectra (like the 40
spectra obtained from 2 RGS detectors in two spectral orders for ten
observations that we have for our Mrk~509 data), this can be a cumbersome
procedure as the memory and CPU requirements become very demanding. Another
option is to rigorously discard all wavelength bins where even during
short periods (a single observation or a part of an observation) data are
lacking. This solution will cause a significant loss of diagnostic capability,
as many more bins are discarded and likely several of those bins will be near
astrophysically interesting features.

Here we follow a different route. It is common wisdom (and this is also
suggested by the SAS manual) not to use fluxed spectra for fitting analysis. The
problem is the lack of a well-defined redistribution function for that case. We
will show later in Sect.~\ref{sect:matrix} that we can obtain an appropriate
response matrix for our case, allowing us to do analysis with fluxed spectra
under the conditions given in that section. 

We first create individual fluxed spectra using the SAS task {\sl rgsfluxer}. We
use exactly the same wavelength grid for each observation. We then average the
fluxed spectra using for each spectral bin the exposure times as weights. This
is done for all bins having 100\% exposure and for all spectra that are to
combined.

For bins with lacking data (like bins A and B of Fig.~\ref{fig:pixel}) we have
to follow a different approach. If the total exposure time of spectrum $k$ is
given by $t_k$, and the effective exposure time of spectral bin $j$ in spectrum
$k$ is given by $t_{kj}$, we clearly have $t_{kj} < t_k$ for ``problematic''
bins $j$, and we will include in the final spectrum only data bins with
$t_{kj}>f t_k$, with $0\le f \le 1$ a tunable parameter discussed below.
However, due to the variability of the source the average flux level  of the
included spectra $k$ for problematic bin $j$ may differ from the flux level of
the  neighbouring bins $j-1$ and $j+1$ that are fully exposed. To correct for
this, we assume that the spectral shape (but not the overall normalisation) in
the local neighbourhood of bin $j$ is constant in time. From these neighbouring
bins (with fluxes $F_{j-1}$ and $F_{j+1}$) we estimate the relative contribution
$R$ to the total flux for the spectra $k$ that have sufficient exposure time
$t_{kj}$ for bin $j$, i.e.
\begin{equation}
R \equiv \frac{\sum\limits_{{k,\ t_{kj}>ft_k}}^{}
t_{k}F_{j-1}} {\sum\limits_{{k}}^{} t_{k}F_{j-1}}, 
\label{eqn:rat}
\end{equation}
and similar for the other neighbour at $j+1$.  We interpolate the values for $R$
linearly at both sides to obtain  a single value at the position of bin $j$. The
flux for bin $j$ is now estimated as 
\begin{equation} 
F_j = \frac{\sum\limits_{{k,\ t_{kj}>ft_k}}^{} t_{k}F_{j}}
{R\ \sum\limits_{{k}}^{} t_{k}}.
\end{equation} 
Example: for bin $A$ in Fig.~\ref{fig:pixel} we have $R=200\times 1/(400\times 1
+ 200\times 1) = 1/3$, and hence using the flux measurement in spectrum 2, we
have $F_A = (200\times 1) / ((1/3)\times 2) = 300$~counts\,s$^{-1}$, as  it
should be.

This procedure gives reliable results as long as the spectral shape does not
change locally; as there is less exposure at bin $j$, the error bar on the flux
will obviously be larger. However, when there is reason to suspect that the bin
is at the location of a spectral line that changes in equivalent width, this
procedure cannot be applied!

\begin{figure}[!htbp]
\resizebox{\hsize}{!}{\includegraphics[angle=-90]{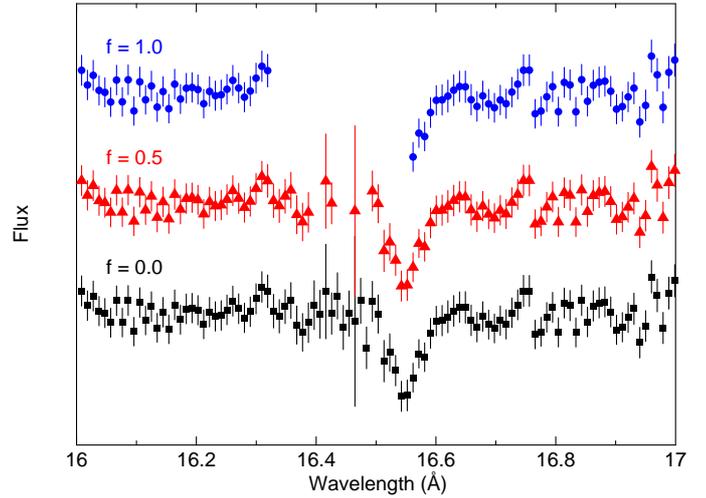}}
\caption{Spectral region near the gap between CCD 5 and 6 of RGS2, containing
the 1s--3p line of \ion{O}{viii} at 16.55~\AA\ from the outflow. From top to
bottom we show the fluxed spectra (arbitrarily scaled and shifted along the flux
axis) for three values of $f$. Using the conservative setting $f=1$, a major
part of the spectrum is lost. With $f=0$ no data are lost but of course between 
16.3 and 16.55~\AA\ the error bars are slightly larger due to the shorter
exposure time in that wavelength region.}
\label{fig:gap}
\end{figure}

To give the user more flexibility, we have introduced the minimum exposure
fraction $f$ in (\ref{eqn:rat}). For $f=0$, we obtain the best results for
spectra with constant shape (as every bit of information that is available is
used). On the other hand, if $f=1$, only those data bins will be included that
have no bad bins in any of the observations to be combined. The advantage in
that case is that there is no bias in the stacked spectrum, but a disadvantage
is that a significant part of the spectrum may be lost, for example near
important diagnostic lines. In particular for the multi-pointing mode the
purpose of which is to have different wavelengths fall on different part of the
spectrum and thus to have a measurement of the spectrum at all wavelengths, this
can be a problem. We illustrate this for the region near the chip gap between
CCD 5 and 6 of RGS2 (see Fig.~\ref{fig:gap}). For $f<1$ some data points have
large error bars due to the low effective exposure of some bins containing
missing columns, and the corresponding low number of counts $N$ yielding large
relative errors ($\sim N^{-0.5}$). In our further analysis we use $f=0.$

Finally, in the combined spectrum due to the binning and randomisation
procedures within the SAS task {\sl rgsfluxer}, it is still possible that
despite careful screening for bad pixels, a few bad bins remain in the  fluxed
spectrum, often adjacent or close to discarded bins. For this reason, we check
for any bins that are more than $3\sigma$ below their right or left neighbouring
bin (taking the statistical errors of both into account). Typically, the
algorithm finds a few additional bad bins in an individual observation, which we
also discard from our analysis. Only for very strong isolated emission lines
with more than 1500 counts in a single observation our method would produce
false rejections near the bending points of the instrumental spectral
redistribution function, because then the spectral changes for neighbouring bins
are stronger than 3$\sigma$ statistical fluctuations, but our spectra of Mrk~509
do not contain such sharp and strong features.

We have implemented the procedures presented in this Section in the program {\sl
rgs\_fluxcombine} that is available within the publicly available SPEX
distribution\footnote{www.sron.nl/spex}.

Finally, we note that in principle another route exists to combine the spectra.
This would be to run {\sl rgscombine} but to modify by a user routine the column
AREASCAL so that the count rates are properly corrected for the missing bins. It
can be shown that in the combined spectrum AREASCAL must be scaled by $1/R$,
with $R$ derived as described above, in order to obtain robust results. It
requires however the development of similar tools as {\sl rgs\_fluxcombine} to
do this modification.

\subsection{Combination of spectra from different RGS or spectral order}

In the previous section we showed how to combine spectra from the same RGS and
spectral order, but for different observations. However, we also want to combine
RGS1 and RGS2 spectra, and spectra from the first and second spectral order. We
do this by simply averaging the fluxed spectra, using for each bin the
statistical errors on the flux as weight factors. Because of the lower effective
area in the second spectral order, second order spectra get in this way less
weight then first order spectra, their weights are typically between 20\% (near
18~\AA) and 50\% (near 8~\AA) of the first-order spectral weights. 

In Sect.~\ref{sect:matrix} we show how to create a response matrix for the
fluxed RGS spectra. The same weights that are used to determine the relative
contributions of the different spectra to the combined spectrum are also used to
weigh the contributions from the corresponding redistribution functions. To
avoid discontinuities near the end points of second order spectra or near
missing CCDs, we pay attention to small effective area corrections, as outlined
in Sect.~\ref{sect:areacor}.

\section{Response matrix\label{sect:matrix}}

We have explained before that for time-variable sources standard SAS tasks like
{\sl rgscombine} cannot be used to combine the 10 individual spectra of Mrk~509
or any other source into a single response matrix. For that reason, we have
combined fluxed spectra as described in the sections above. The only other
feasible alternative would be to fit the 10 individual spectra simultaneously.
But with 2 RGS detectors, 2 spectral orders and 10 observations, this adds up to
40 individual spectra. The memory occupied by the corresponding response
matrices is 2.0 Gbyte. Although our fitting program SPEX is able to cope with
this, fitting becomes cumbersome and error searches extremely slow, due to the
large number of matrix multiplications that have to be performed.

For our combined fluxed spectra, fitting is much faster because we have only a
single spectrum and hence only need one response matrix. As we use fluxed
spectra, we can use a simple response matrix with unity effective area, and the
spectral redistribution function given by the redistribution part of the RGS
matrix. Unfortunately, the RGS response matrix produced by the SAS {\sl
rgsrmfgen} task combines the effective area and redistribution part into a
single data file. Furthermore, due to the multi-pointing mode that we used and
due to transient bad columns, the matrix for each of the 10 observations will be
slightly different.

Therefore, we have adopted the following approach. For a number of wavelengths
(7 to 37~\AA, step size 2~\AA), we have fitted the RGS response to the sum of
three or four Gaussians. This gives a quasi-diagonal matrix resulting in another
speed-up of spectral analysis. For these fits to the redistribution function, we
have omitted the data channels with incomplete exposure (near chip boundaries,
and at bad pixels). The parameters of the Gaussians (normalisation, centre and
width) were then modelled with smooth analytical functions of wavelength. We
paid most attention to the peak of the redistribution function, where most
counts are found.

\begin{table}[!ht]
\caption[]{Parameters of the RGS redistribution function.}
\smallskip
\label{tab:redispar}
\begin{tabular}{lrrrr}
\hline
RGS & RGS1 & RGS2 & RGS1 & RGS2 \\
Order & $-1$ & $-1$ & $-2$ & $-2$ \\
\hline
$a_1$ & 0.0211    & 0.0237    &   0.0100  & 0.0118 \\
$a_2$ & 0.0514    & 0.055     &   0.018   & 0.027  \\
$a_3$ & 0.0105    & 0.016     &$-$0.0572  & 0.017  \\
$a_4$ &   --      &  --       &   0.404   & 0.339  \\
$b_1$ & 0.00028   & 0.00032   &   0.00031 & 0.00035 \\
$b_2$ & 0.00039   & 0.00058   &   0.00075 & 0.00080 \\
$b_3$ & 0.021     & 0.020     &   0.021   & 0.00066 \\
$b_4$ &   --      &  --       &$-$0.01    & --     \\
$c_3$ &$-$0.000405&$-$0.000346&$-$0.000692& --     \\
$d_2$ & 0.1068    & 0.1314    & $-$1.1022 & 0.0084 \\
$d_3$ & --        & --        &1.5276     &  1.2276 \\
$d_4$ & --        & --        &   0.3190  &$-$0.1211 \\
$e_2$ & 0.0125    & 0.0086    & 0.289     & 0.0153 \\
$e_3$ & --        & --        &$-$0.223   &$-$0.140 \\
$e_4$ & --        & --        &$-$0.017   &   0.0833 \\
$f_2$ &$-$0.00024 &$-$0.00017 &$-$0.0215  &$-$0.00024 \\
$f_3$ & --        & --        & 0.016     &  0.0075 \\
$f_4$ & --        & --        &$-$0.0014  &$-$0.0080 \\
$g_2$ & --        & --        & 0.00053   &$-$0.000009 \\
$g_3$ & --        & --        &$-$0.00044 &$-$0.00015 \\
$g_4$ & --        & --        &   0.00011 &   0.00022 \\
\hline
\end{tabular}
\end{table}

The following parametrisation describes the RGS redistribution functions well
(all units are in \AA):
\begin{equation}
R(\lambda^{\prime},\lambda) = \sum\limits_{i=1}^4 \frac{N_i}{\sqrt{2\pi}
\sigma_i} \mathrm{e}^{\displaystyle{-(\lambda-\lambda^{\prime})^2/2\sigma_i^2}},
\end{equation}
\begin{equation}
\sigma_i = a_i + b_i\lambda + c_i \lambda^2,
\end{equation}
\begin{equation}
i=1:\ \ N_1 = 1 - N_2 - N_3 - N_4,
\end{equation}
\begin{equation}
i=2,\ {\rm 1st\ order\ only:}\ \ N_i = d_i + e_i\lambda + f_i \lambda^2 + g_i\lambda^3,
\end{equation}
\begin{equation}
i=2-4,\ {\rm 2nd\ order\ only:}\ \ N_i = d_i + e_i\lambda + f_i \lambda^2 + g_i\lambda^3,
\end{equation}
\begin{equation}
i=3,\ {\rm 1st\ order\ RGS1\ only:}\ \ N_3 = -0.065 + 2.5 / \lambda^{0.7},
\end{equation}
\begin{equation}
i=3,\ {\rm 1st\ order\ RGS2\ only:}\ \ N_3 = -0.075 + 2.5 / \lambda^{0.7}.
\end{equation}

\begin{figure}
\resizebox{\hsize}{!}{\includegraphics[angle=-90]{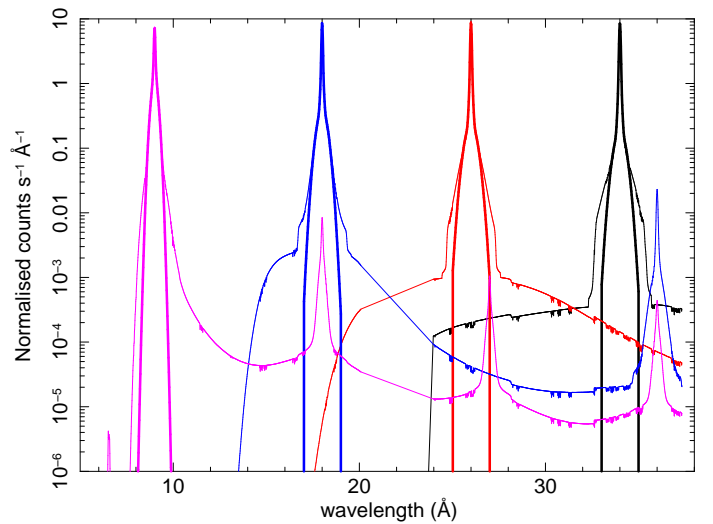}}
\caption{Comparison of the redistribution function for RGS2, first order
as delivered by the SAS (thin solid line) to our approximation using three
Gaussians (thick lines), for four different photon wavelengths.}
\label{fig:redis1}
\end{figure}

\begin{figure}
\resizebox{\hsize}{!}{\includegraphics[angle=-90]{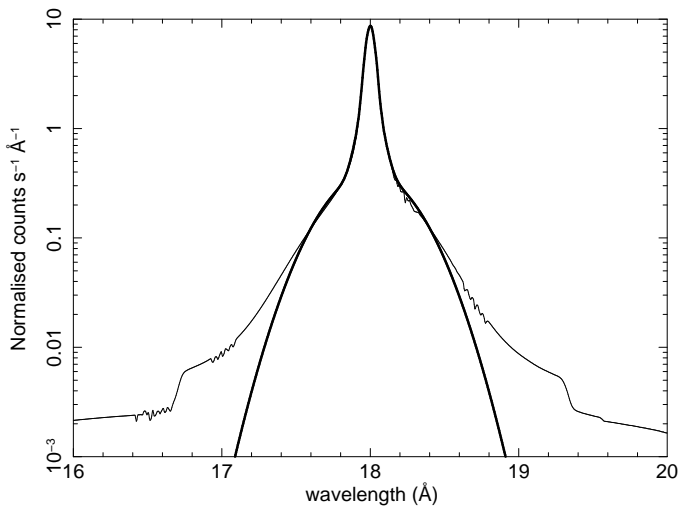}}
\caption{As Fig.~\ref{fig:redis1}, but only for photons at 18~\AA.}
\label{fig:redis2}
\end{figure}

The relevant parameters are given in Table~\ref{tab:redispar}, omitting
parameters that are not used (zero). We cut-off the redistribution function
beyond $\pm 1$~\AA\ from the line centre. We have compared these approximations
to the true redistribution function (see
Figs.~\ref{fig:redis1}--\ref{fig:redis2}), and find that our model accurately
describes the core down to a level of about 1~\% of the peak of the
redistribution function. Also, the flux outside the $\pm 1$~\AA\  band is less
than $\sim 1$\% of the total flux for all wavelengths. It should be noted that
the above approximation is more accurate than the calibration of the
redistribution itself. Currently, the width of the redistribution is known only
to about 10\% accuracy\footnote{See also \citet{kaastra2006}, Table~2 where
equivalent widths based on different representations of the redistribution
function are compared, and also a comparison to Chandra LETGS data is made.}. 

We finally remark that Fig.~\ref{fig:redis1} seems to suggest that the response
also contains the second order spectra, for instance for 9~\AA\ photons, apart
from the main peak near 9~\AA, there are secondary peaks at 18, 27 and 36~\AA. 
These peaks are, however, not the genuine higher order spectra, but they
correspond to higher order photons that by chance end up in the low-energy tail
of the CCD redistribution function. For instance, a photon with intrinsic energy
1.38~keV would be seen in first order at 9~\AA, but in second order at 18~\AA. A
small fraction of these second order events end up in the CCD redistribution
tail and are not detected near 1.38~keV but near 0.69~keV. Because their (second
order) measured  wavelength is 18~\AA, they are identified as first order events
(energy 0.69~keV, wavelength 18~\AA). As can be seen from Fig.~\ref{fig:redis1},
this is only a small contribution, and we safely can ignore them.

\section{Effective area corrections\label{sect:areacor}}

\begin{figure}[!htbp]
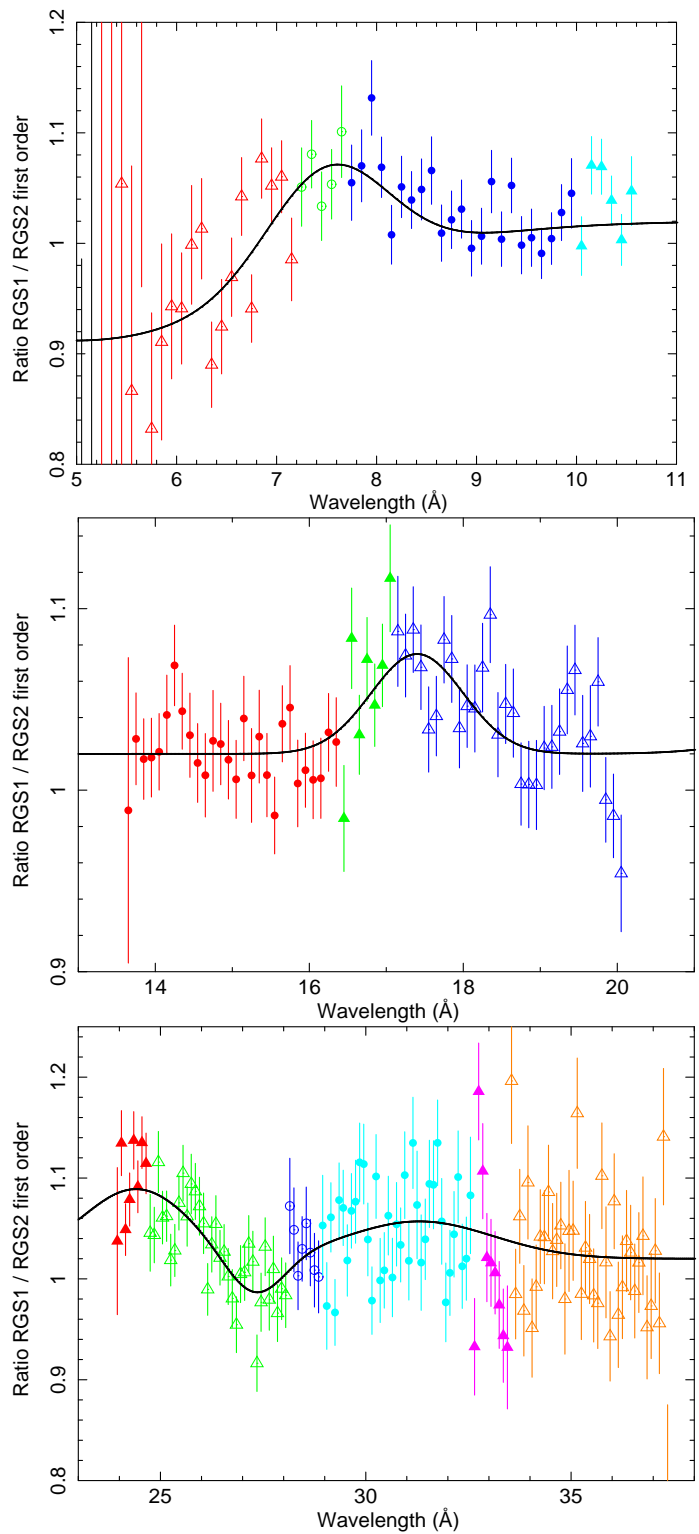

\resizebox{\hsize}{!}{\includegraphics[angle=-90]{ratio11to21a.cps}}
\resizebox{\hsize}{!}{\includegraphics[angle=-90]{ratio11to21b.cps}}
\resizebox{\hsize}{!}{\includegraphics[angle=-90]{ratio11to21c.cps}}
\caption{Flux ratio between RGS1 and RGS2 for parts of the spectrum that are
detected by both RGS. Different symbols
indicate different combinations of CCD chips for both RGS detectors. 
Open symbols: RGS1 chips 1, 3, 5, 9; filled symbols: RGS1 chips 2, 4, 6, 8.
Triangles: RGS2 chips 1, 3, 5, 7, 9; circles: RGS2 chips 2, 6, 8. The solid
line is a simple fit to the ratio's using a constant plus Gaussians.}
\label{fig:ratio1}
\end{figure}

\begin{figure}[!htbp]
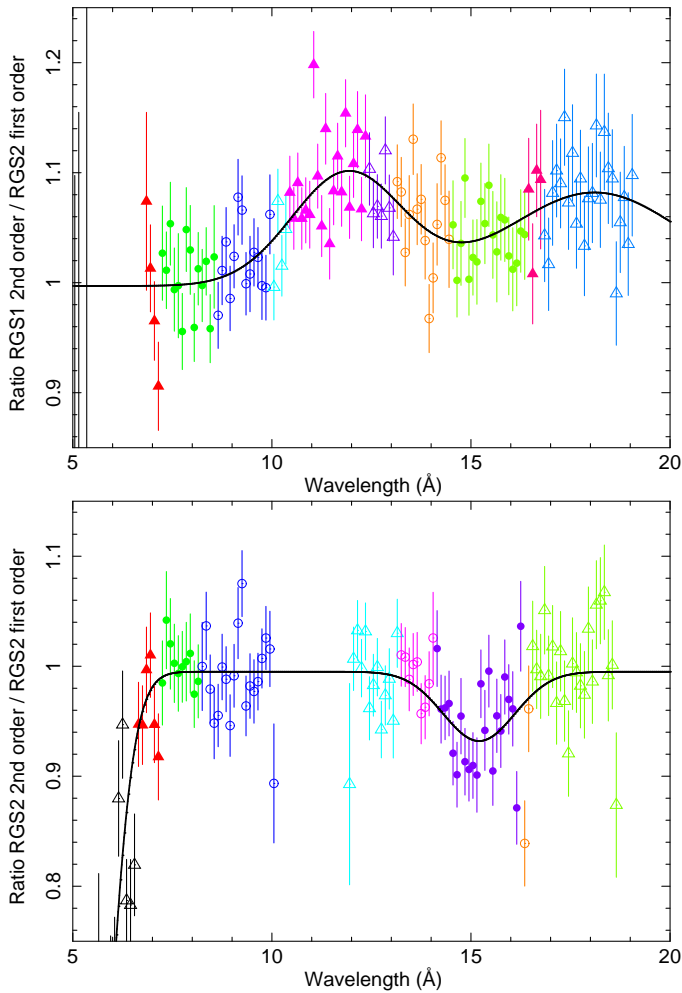

\resizebox{\hsize}{!}{\includegraphics[angle=-90]{ratio12to21.cps}}
\resizebox{\hsize}{!}{\includegraphics[angle=-90]{ratio22to21.cps}}
\caption{As Fig.~\ref{fig:ratio1}, but for second order spectra compared to
first order spectra. Due to the specific geometry of missing CCDs (CCD7 for
RGS1, CCD4 for RGS2) we compare all fluxes to the first order of RGS2. Open 
symbols: second order chips 1, 3, 5, 7, 9; filled symbols: second order chips 2,
4, 6, 8. Triangles: first order chips 1, 3, 5, 7, 9; circles: first order chips
2, 6, 8.
}
\label{fig:ratio2}
\end{figure}

The statistics of our stacked 600~ks spectrum of Mrk~509 is excellent, with
statistical uncertainties down to 2\% per 0.1~\AA\ per RGS in first order. In
larger bins of 1~\AA\ wide, the quality is even better by a factor of $\sim 3$.
These statistical uncertainties are smaller than the accuracy to which the RGS
effective area has been calibrated. A close inspection of the first and second
order spectra of our source for both RGS detectors showed that there are some
systematic flux differences of up to a few percent at wavelength scales larger
than a few tenths of an \AA\ (Figs.~\ref{fig:ratio1}--\ref{fig:ratio2}),
consistent with the known accuracy of the effective area calibration. These
differences sometimes correlate with chip boundaries (like at 34.5~\AA) but not
always. A preliminary analysis shows the same trends for stacked RGS spectra of
the blazar Mrk~421.

While a full analysis is underway, we use here a simple approach to correct
these differences and obtain an accurate relative effective area calibration. We
model the differences purely empirical with the sum of a constant function and a
few Gaussians with different centroids, widths and amplitudes, as indicated in
Figs.~\ref{fig:ratio1}--\ref{fig:ratio2}. We then attribute half of the
deviations to RGS1 and the other half to RGS2. In this way there is better
agreement between the different RGS spectra, in particular across spectral
regions where due to the failure of one of the chips the flux would otherwise
show a discontinuity at the boundary between a region with two chips and a
region with only one chip. As the effective area of the second order spectra has
been calibrated with less accuracy than the first order spectra, we adjust the
second order spectrum to match the first order spectrum.

\begin{figure}
\resizebox{\hsize}{!}{\includegraphics[angle=-90]{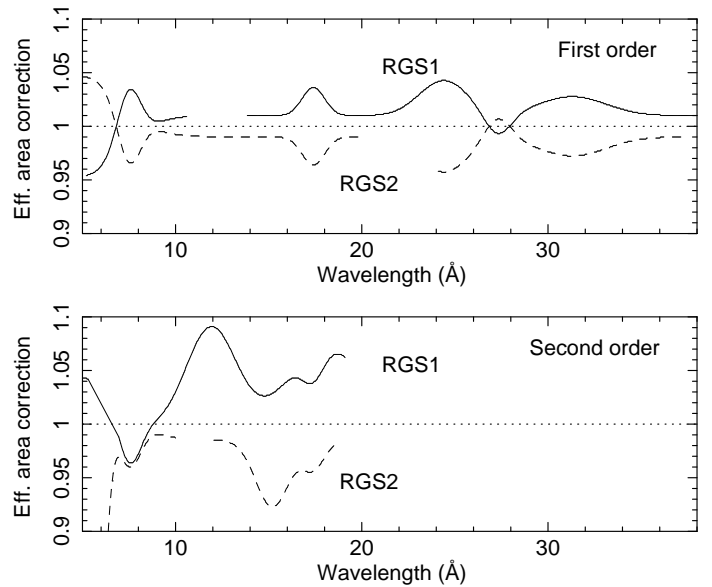}}
\caption{Adopted effective area corrections for RGS1 (solid lines) and RGS2
(dashed lines), in first and second spectral order.}
\label{fig:areacor}
\end{figure}

We obtain the following corrections factors:
\begin{eqnarray}
r_{11}(\lambda) = 1.02 &+& G(\lambda,-0.108,5,2) \nonumber \\
              &+& G(\lambda,0.099,7.5,0.6) \nonumber \\
              &+& G(\lambda,0.055,17.4,0.6) \nonumber \\
	      &+& G(\lambda,0.069,24.4,1.3) \nonumber \\
              &+& G(\lambda,-0.042,27.3,0.7) \nonumber \\
	      &+& G(\lambda,0.037,31.3,1.8), \\
r_{12}(\lambda) = 0.997 &+& G(\lambda,0.103,11.9,1.4)  \nonumber \\
               &+& G(\lambda,0.085,18.1,2.2), \\
r_{22}(\lambda) = 0.995 &+& G(\lambda,-0.882,5.0,0.7)  \nonumber \\
               &+& G(\lambda,-0.063,15.2,0.9),
\end{eqnarray}
Here $r_{ij}$ is the flux of RGS $i$ order $j$ relative to RGS2 order 1 as shown
in Figs.~\ref{fig:ratio1}--\ref{fig:ratio2}, and $G(\lambda,N,\mu,\sigma)\equiv
N {\rm e}^{\displaystyle{-(\lambda-\mu)^2/2\sigma^2}}$. With these definitions, the
fluxes need to be multiplied by correction factors $a_{ij}$ to get the corrected
values as follows:
\begin{eqnarray}
a_{11} &=& (1 + 1/r_{11}) / 2 \\
a_{21} &=& (1 + r_{11}) / 2  \\
a_{12} &=& (1 + r_{11}) / 2r_{12}  \\
a_{22} &=& (1 + r_{11}) / 2r_{22}.
\end{eqnarray}

The above approach may fail whenever the small discrepancy between one spectrum
and the other is caused by a single RGS, and not by both. Then there will be a
systematic error in the derived flux with the same Gaussian shape as used for
the correction factors, but with half its amplitude. This is compensated
largely, however, by our approach where we fit the true underlying continuum of
Mrk~509 by a spline, and in our analysis we make sure that we do not attribute
real broad lines to potential effective area corrections. The adopted effective
area corrections (the inverse of the $a_{ij}$ factors) are shown in
Fig.~\ref{fig:areacor}. Note that for wavelength ranges with lacking CCDs the
effective area corrections are based on the formal continuation of the above
equations, and because of that may differ from unity. However, this is still
consistent with the absolute accuracy of the RGS effective area calibration.

\section{Wavelength scale}

\subsection{Aligning both RGS detectors\label{sect:aligning}}

The zero-point of the RGS wavelength scale has a systematic uncertainty of about
8 m\AA\ \citep{denherder2001}. According to the latest insights, this is
essentially due to a slight tilting of the grating box due to Solar irradiance
on one side of it, which causes small temperature differences and hence
inhomogeneous thermal expansion. The thermal expansion changes the incidence
angle of the radiation on the gratings and hence the  apparent wavelength. The
details of this effect are currently under investigation. Due to the varying
angle of the satellite with respect to the Sun during an orbit and the finite
thermal conductivity time scale of the relevant parts, the precise correction
will depend also on the history of the satellite orbit and pointing.

\begin{figure}
\resizebox{\hsize}{!}{\includegraphics[angle=-90]{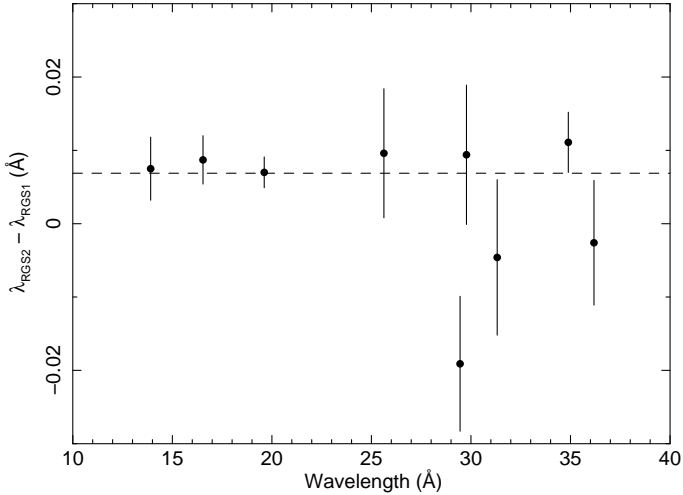}}
\caption{Wavelength difference RGS2 -- RGS1 for the 9 strongest lines. The dashed
line indicates the weighted average of $6.9\pm 0.7$~m\AA.}
\label{fig:wrel}
\end{figure}

As corrections for this effect are still under investigation, we take here an
empirical approach. We start with aligning both RGS detectors. We have taken the
9 strongest spectral lines in our Mrk~509 spectrum that can be detected on both
RGS detectors, and have determined the difference between the average measured
wavelength for both RGS detectors in the combined spectrum. We found that the
wavelengths measured with RGS2 are $6.9\pm 0.7$~m\AA\ larger than the
wavelengths measured with RGS1 (Fig.~\ref{fig:wrel}).

Therefore, we have re-created the RGS2 spectra by sampling them on a wavelength
grid similar to the RGS1 wavelength grid, but shifted by $+6.9$~m\AA. Using this
procedure, spectral lines always end up in the same spectral bins. This allows
us to stack RGS1 and RGS2 spectra by simply adding up the counts for each
spectral bin, without a further need to re-bin the counts when combining both
spectra.

After applying this correction, we stacked the spectra of both RGS detectors and
compared the first and second order spectra. We found that the second order
centroids agree well with the first order centroids (to within an accuracy of
3~m\AA) and therefore did not apply a correction.

\subsection{Strongest absorption lines}

\begin{figure*}[!htbp]
\resizebox{\hsize}{!}{\includegraphics[angle=-90]{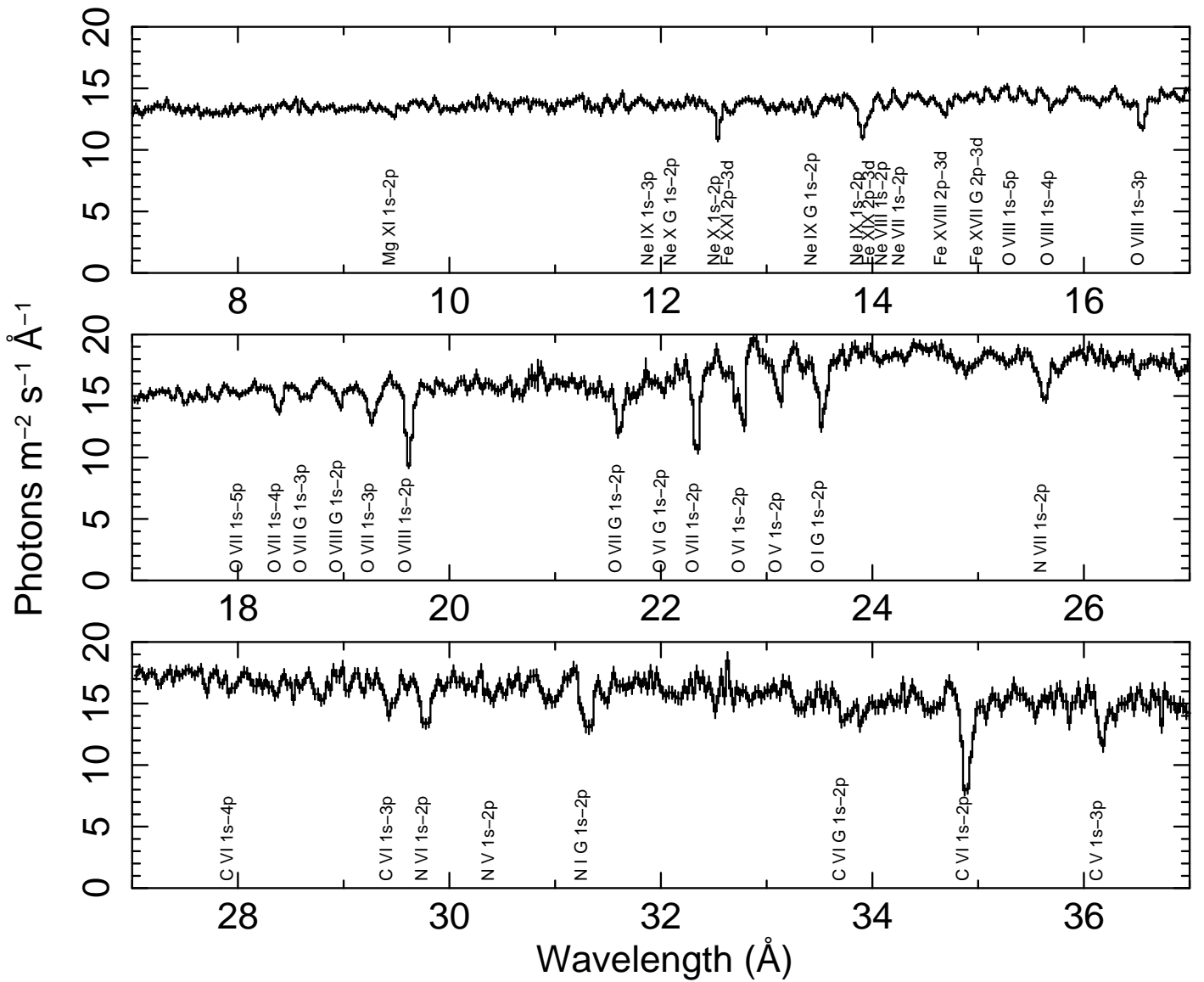}}
\caption{Fluxed, stacked spectrum of Mrk~509. Only the lines used in
Table~\ref{tab:lines} have been indicated. See \citet{detmers2011} for a
more detailed figure.}
\label{fig:spectrum}
\end{figure*}

\begin{table*}[!ht]
\caption[]{Strongest absorption lines observed in the spectrum of Mrk~509.}
\smallskip
\label{tab:lines}
\begin{tabular}{llccccc}
\hline
Ion & Trans-- & $EW$$^{\mathrm{a}}$ & $\lambda_{\mathrm {obs}}$$^{\mathrm{b}}$ & 
   $\lambda_{\mathrm {lab}}$$^{\mathrm{c}}$& $v$$^{\mathrm{d}}$ & Reference$^{\mathrm{e}}$ \\
    &  ition  & (m\AA) & \AA\ & \AA\ & (km\,s$^{-1}$) & \\
\hline
\ion{Fe}{xxi}   & 2p--3d & $ 3.0\pm 0.8$ & $12.664\pm 0.008$ & $12.2840\pm 0.0020$ & $ -990\pm 200$ &  \citet{brown2002} \\ 
\ion{Fe}{xix}   & 2p--3d & $ 8.5\pm 1.2$ & $13.979\pm 0.005$ & $13.516 \pm 0.005 $ & $  -20\pm 150$ &  \citet{phillips1999} \\
\ion{Fe}{xviii} & 2p--3d & $ 5.6\pm 0.8$ & $14.680\pm 0.005$ & $14.207 \pm 0.010 $ & $ -330\pm 230$ &  \citet{phillips1999} \\ 
\ion{Mg}{xi}    & 1s--2p & $ 3.6\pm 0.7$ & $ 9.465\pm 0.005$ & $ 9.1688\pm 0.0000$ & $ -590\pm 150$ &  \citet{flemberg1942} \\ 
\ion{Ne}{x}     & 1s--2p & $ 9.8\pm 0.9$ & $12.539\pm 0.002$ & $12.1346\pm 0.0000$ & $ -300\pm  40$ &  \citet{erickson1977} \\
\ion{Ne}{ix}    & 1s--3p & $ 2.5\pm 0.9$ & $11.921\pm 0.009$ & $11.5440\pm 0.0020$ & $ -480\pm 240$ &  \citet{peacock1969} \\ 
\ion{Ne}{ix}    & 1s--2p & $13.2\pm 1.1$ & $13.897\pm 0.003$ & $13.4470\pm 0.0020$ & $ -260\pm  70$ &  \citet{peacock1969} \\ 
\ion{Ne}{viii}  & 1s--2p & $ 5.6\pm 0.8$ & $14.118\pm 0.006$ & $13.654 \pm 0.005 $ & $ -100\pm 160$ &  \citet{peacock1969} \\ 
\ion{Ne}{vii}   & 1s--2p & $ 4.0\pm 0.8$ & $14.282\pm 0.006$ & $13.8140\pm 0.0010$ & $ -130\pm 130$ &  \citet{behar2002} \\ 
\ion{O}{viii}   & 1s--5p & $ 2.4\pm 0.9$ & $15.332\pm 0.010$ & $14.8205\pm 0.0000$ & $  +30\pm 190$ &  \citet{erickson1977} \\
\ion{O}{viii}   & 1s--4p & $ 5.2\pm 0.8$ & $15.692\pm 0.006$ & $15.1762\pm 0.0000$ & $ -120\pm 110$ &  \citet{erickson1977} \\
\ion{O}{viii}   & 1s--3p & $13.4\pm 1.2$ & $16.546\pm 0.002$ & $16.0059\pm 0.0000$ & $ -190\pm  40$ &  \citet{erickson1977} \\
\ion{O}{viii}   & 1s--2p & $31.8\pm 1.4$ & $19.615\pm 0.001$ & $18.9689\pm 0.0000$ & $ -110\pm  20$ &  \citet{erickson1977} \\
\ion{O}{vii}    & 1s--5p & $ 2.7\pm 0.9$ & $18.022\pm 0.013$ & $17.3960\pm 0.0020$ & $ +460\pm 230$ &  \citet{engstrom1995} \\ 
\ion{O}{vii}    & 1s--4p & $10.3\pm 1.3$ & $18.382\pm 0.004$ & $17.7683\pm 0.0007$ & $  +40\pm  60$ &  \citet{engstrom1995} \\ 
\ion{O}{vii}    & 1s--3p & $14.8\pm 1.0$ & $19.265\pm 0.003$ & $18.6284\pm 0.0004$ & $  -70\pm  50$ &  \citet{engstrom1995} \\ 
\ion{O}{vii}    & 1s--2p & $32.7\pm 2.1$ & $22.335\pm 0.002$ & $21.6019\pm 0.0003$ & $ -140\pm  20$ &  \citet{engstrom1995} \\ 
\ion{O}{vi}     & 1s--2p & $21.2\pm 2.3$ & $22.771\pm 0.005$ & $22.0194\pm 0.0016$ & $  -90\pm  70$ &  \citet{schmidt2004} \\ 
\ion{O}{v}      & 1s--2p & $15.8\pm 2.9$ & $23.120\pm 0.005$ & $22.3700\pm 0.0100$ & $ -260\pm 140$ &  \citet{gu2005} \\ 
\ion{N}{vii}    & 1s--2p & $19.0\pm 2.3$ & $25.625\pm 0.004$ & $24.7810\pm 0.0000$ & $ -110\pm  50$ &  \citet{erickson1977} \\
\ion{N}{vi}     & 1s--2p & $23.8\pm 2.8$ & $29.773\pm 0.004$ & $28.7875\pm 0.0002$ & $  -70\pm  40$ &  \citet{engstrom1995} \\ 
\ion{N}{v}      & 1s--2p & $ 7.6\pm 2.3$ & $30.402\pm 0.014$ & $29.414 \pm 0.004 $ & $ -240\pm 150$ &  \citet{beiersdorfer1999} \\ 
\ion{C}{vi}     & 1s--4p & $ 5.7\pm 1.9$ & $27.933\pm 0.015$ & $26.9898\pm 0.0000$ & $ +150\pm 160$ &  \citet{erickson1977} \\
\ion{C}{vi}     & 1s--3p & $16.2\pm 2.8$ & $29.438\pm 0.007$ & $28.4656\pm 0.0000$ & $  -90\pm  70$ &  \citet{erickson1977} \\
\ion{C}{vi}     & 1s--2p & $49.3\pm 3.0$ & $34.890\pm 0.002$ & $33.7360\pm 0.0000$ & $  -70\pm  20$ &  \citet{erickson1977} \\
\ion{C}{v}      & 1s--3p & $17.1\pm 3.2$ & $36.174\pm 0.005$ & $34.9728\pm 0.0008$ & $  -30\pm  40$ &  \citet{edlen1970} \\ 
\hline
\ion{Fe}{xvii} G& 2p--3d & $ 2.5\pm 0.8$ & $15.021\pm 0.008$ & $15.0140\pm 0.0010$ & $ +160\pm 160$ &  \citet{brown1998} \\ 
\ion{Ne}{x}    G& 1s--2p & $ 1.6\pm 0.8$ & $12.125\pm 0.010$ & $12.1346\pm 0.0000$ & $ -210\pm 240$ &  \citet{erickson1977} \\
\ion{Ne}{ix}   G& 1s--2p & $ 4.2\pm 0.9$ & $13.454\pm 0.006$ & $13.4470\pm 0.0020$ & $ +180\pm 150$ &  \citet{peacock1969} \\
\ion{O}{viii}  G& 1s--2p & $ 8.2\pm 1.1$ & $18.966\pm 0.005$ & $18.9689\pm 0.0000$ & $  -30\pm  80$ &  \citet{erickson1977} \\
\ion{O}{vii}   G& 1s--3p & $ 5.7\pm 1.1$ & $18.624\pm 0.010$ & $18.6284\pm 0.0004$ & $  -70\pm 170$ &  \citet{engstrom1995} \\ 
\ion{O}{vii}   G& 1s--2p & $16.6\pm 2.2$ & $21.604\pm 0.004$ & $21.6019\pm 0.0003$ & $  +30\pm  50$ &  \citet{engstrom1995} \\ 
\ion{O}{vi}    G& 1s--2p & $ 4.5\pm 1.8$ & $22.025\pm 0.013$ & $22.0194\pm 0.0016$ & $  +70\pm 180$ &  \citet{schmidt2004} \\ 
\ion{C}{vi}    G& 1s--2p & $ 8.8\pm 2.5$ & $33.743\pm 0.013$ & $33.7360\pm 0.0000$ & $  +80\pm 170$ &  \citet{erickson1977} \\
\hline
\ion{O}{i}     G& 1s--2p & $29.1\pm 2.8$ & $23.521\pm 0.003$ & $23.5113\pm 0.0018$ & $ +130\pm  50$ &  \citet{kaastra2010} \\
\ion{N}{i}     G& 1s--2p & $22.2\pm 4.3$ & $31.302\pm 0.009$ & $31.2857\pm 0.0005$ & $ +160\pm  80$ &  \citet{santanna2000} \\
\hline
\end{tabular}
\smallskip
\begin{list}{}{}
\item[$^{\mathrm{a}}$] Equivalent width
\item[$^{\mathrm{b}}$] Observed wavelength using the nominal SAS wavelength scale for RGS1, with RGS2 wavelengths shifted
downwards by 6.9~m\AA.
\item[$^{\mathrm{c}}$] Laboratory wavelength (experimental or theoretical) with best estimate of uncertainty
\item[$^{\mathrm{d}}$] Outflow velocity; for lines from the AGN relative to the
systematic redshift of 0.034397; for lines from the ISM (labelled ``G'' in the
first column) to the LSR velocity scale. Observed errors and errors in the
laboratory wavelength have been combined. The velocities are based on the
observed wavelengths of column (4) but corrected for the orbital motion of the
Earth, and have in addition been adjusted by adding 1.8~m\AA\ to the observed
wavelengths (see Sect.~\ref{sect:wavsummary}).
\item[$^{\mathrm{e}}$] References for the laboratory wavelengths
\end{list}
\end{table*}

To facilitate the derivation of the proper wavelength scale, we have compiled a
list of the strongest spectral lines that are present in the spectrum of Mrk~509
(Table~\ref{tab:lines}). We have measured the centroids $\lambda_0$ and
equivalent widths $EW$ of these lines by fitting the fluxed spectra $F(\lambda)$
locally in a 1~\AA\ wide band around the line centre using Gaussian absorption
lines, superimposed on a smooth continuum for which we adopted a quadratic
function of wavelength. We took care to re-formulate this model in such a way
that the equivalent width is a free parameter of the model and not a derived
quantity based on the measured continuum or line peak:
\begin{equation}
F(\lambda ) =  c \Bigl\{ 1 + b(\lambda-\lambda_0)+a(\lambda-\lambda_0)^2
   - \frac{EW}{\sigma\sqrt{2\pi}}
   {\mathrm{e}}^{-(\lambda-\lambda_0)^2/2\sigma^2)}
   \Bigr\} .
\end{equation}

As we focus in this paper entirely on absorption lines, we report here $EW$s for
absorption lines as positive numbers. Where needed we added additional Gaussians
to model the other strong but narrow absorption or emission lines that are
present in the same 1~\AA\ wide intervals around the line of interest. The
Gaussians include the instrumental broadening. For the strongest 14 lines we
could leave the width of the Gaussians a free parameter; for the others we
constrained the width to a smooth interpolation of the width from these stronger
lines.

For illustration purposes, we show the full fluxed spectrum (after all
corrections that we derived in this paper have ben applied) in
Fig.~\ref{fig:spectrum}. A more detailed view of the spectrum including
realistic spectral fits is given in paper III of this series
\citep{detmers2011}.

\subsection{Doppler shifts and velocity scale}

The spectra as delivered by the SAS are not corrected for any Doppler shifts. We
have estimated that the time-averaged velocity of the Earth with respect to
Mrk~509 is $-29.3$~km\,s$^{-1}$ for the present Mrk~509 data. Because the angle
between the lines of sight towards the Sun and Mrk~509 is close to 90\degr\
during all our observations, the differences in Doppler shift between the
individual observations are always less than 1.0~km\,s$^{-1}$, with an r.m.s.
variation of 0.5~km\,s$^{-1}$. Therefore these differences can be ignored (less
than 0.1~m\AA\ for all lines). The orbital velocity of XMM-Newton with respect
to Earth is less than 1.2~km\,s$^{-1}$ for all our observations and can also be
neglected. 

We need a reference frame to determine the outflow velocities of the AGN. For
the reference cosmological redshift we choose here $0.034397\pm 0.000040$
\citep{huchra1993}, as this value is also recommended by the NASA/IPAC
Extragalactic database NED, although the precise origin of this number is
unclear; \citet{huchra1993} refer to a private communication to Huchra et al. in
1988.

Because we do not correct the RGS spectra for the orbital motion of the Earth
around the Sun, we will instead use an effective redshift of $0.03450\pm
0.00004$ to calculate predicted wavelengths. This effective redshift is simply
the combination of the 0.034397 cosmological value with the average
$-29.3$~km\,s$^{-1}$ of the motion of the Earth away from Mrk~509.

For Galactic foreground lines it is common practice to use velocities in the
Local Standard of Rest (LSR) frame, and we will adhere to that convention for
Galactic lines. The conversion from LSR to Heliocentric velocities for Mrk~509
is given by $v_{\mathrm{LSR}} = v_{\mathrm{Helio}} +
8.9\,{\mathrm{km}}\,{\mathrm s}^{-1}$, cf. \citet{sembach1999}.

\subsection{Absolute wavelength scale}

The correction that we derived in Sect.~\ref{sect:aligning} aligns both RGS
detectors and spectral orders, but there is still an uncertainty (wavelength
offset) in the absolute wavelength scale of RGS1. We estimate here this offset
by comparing measured line centroids in our spectrum to predicted wavelengths
based on other X-ray, UV or optical spectroscopy of Mrk~509. We consider here
five different methods:

\begin{enumerate}
\item Lines from the outflow of Mrk~509
\item Lines from foreground neutral gas absorption
\item Lines from foreground hot gas absorption
\item Comparison with a sample of RGS spectra of stars
\item Comparison with Chandra LETGS spectra
\end{enumerate}

\subsubsection{Lines from the outflow of Mrk~509\label{sect:outflowlines}}

The outflow of Mrk~509 has been studied in the past using high-resolution UV
spectroscopy with FUSE \citep{kriss2000} and HST/STIS \citep{kraemer2003}. A
preliminary study of the COS spectra taken simultaneously with the Chandra LETGS
spectra in December 2009, within a month from the present XMM-Newton
observations, shows that there are no major changes in the structure of the
outflow as seen through the UV lines. Hence, we take the archival FUSE and STIS
data as templates for the X-ray lines. \citet{kriss2000} showed the presence of
seven velocity components labelled 1--7 from high to low outflow velocity.
Component 4 may have sub-structure \citep{kraemer2003} but we ignore here these
small differences. The outflow components form two distinct troughs, a high
velocity component from the overlapping components 1--3 and a low velocity
component from the overlapping components 4--7. 

\begin{table}[!ht]
\caption[]{Outflow velocities in km\,s$^{-1}$ derived from UV lines.}
\smallskip
\label{tab:uvoutflow}
\begin{tabular}{lcrrrll}
\hline
Feature & $\log\xi$$^{\mathrm{a}}$ & Comp       & Comp       & Comp       &
     w$^{\mathrm{b}}$ & Ref$^{\mathrm{c}}$\\
        &                          &       1--3 &       4--7 &       1--7 && \\
\hline
\ion{H}{i} $\lambda 1026    $ &        & $-317$ & $ +6$ &   $-56$ & N & FUSE \\
\ion{H}{i} $\lambda 1026    $ &        & $-330$ & $+48$ &   $-89$ & EW& FUSE \\
\ion{O}{vi} $\lambda 1032/38$ & $+0.4$ & $-329$ & $+44$ &   $-28$ & N & FUSE \\
\ion{O}{vi} $\lambda 1032/38$ &        & $-338$ & $+64$ &   $-78$ & EW& FUSE \\
\ion{N}{v} $\lambda 1239/43 $ & $-0.1$ & $-344$ & $+21$ &  $-166$ & N & STIS \\
\ion{C}{iv}$\lambda 1548/51 $ & $-0.6$ & $-328$ & $ +1$ &  $-197$ & N & STIS \\
\ion{C}{iii} $\lambda 977.02$ & $-1.4$ & $-347$ & $-16$ &  $-224$ & N & FUSE \\
\ion{C}{iii} $\lambda 977.02$ &        & $-339$ & $-27$ &  $-231$ & EW& FUSE \\
\hline
\end{tabular}
\smallskip
\begin{list}{}{}
\item[$^{\mathrm{a}}$] ionisation parameter in $10^{-9}$~W\,m
\item[$^{\mathrm{b}}$] N: weighted average using column densities;
                       EW: weighted average using equivalent widths
\item[$^{\mathrm{c}}$] FUSE: \citet{kriss2000}, STIS: \citet{kraemer2003} 
\end{list}
\end{table}

Note that our analysis of ISM lines in the HST/COS spectra for our campaign
\citep{kriss2011} shows that the wavelength scale for the FUSE data
\citep{kriss2000} needs a slight adjustment. As a consequence, we subtract
26~km\,s$^{-1}$ from the velocities reported by \citet{kriss2000}, in addition
to the correction to a different host galaxy redshift scale. 

In Table~\ref{tab:uvoutflow} we list the column density weighted or equivalent
width weighted average velocities. We also give the typical ionisation parameter
$\xi$ for which the ion has its peak concentration, based upon the analysis of
archival XMM-Newton data of Mrk~509 by \citet{detmers2010}.

In principle, equivalent-width weighted line centroids are preferred to predict
the centroids of the unresolved X-ray lines. As for most of the X-ray lines the
optical depths are not very large, using just the column density weighted
average will not make much difference, however.

\begin{table}[!ht]
\caption[]{Wavelength differences (observed minus predicted, in m\AA) for
selected X-ray lines of the outflow.}
\smallskip
\label{tab:xoutflow}
\begin{tabular}{lcccc}
\hline
Line & $\log\xi^{\mathrm{a}}$ & $v^{\mathrm{b}}$ & $\Delta v^{\mathrm{c}}$ &
$\lambda_{\mathrm{obs}} - \lambda_{\mathrm{pred}}$$^{\mathrm{d}}$ \\
 & & (km\,s$^{-1}$) & (km\,s$^{-1}$) & m\AA \\
\hline
\ion{N}{vi} $\lambda 28.79$ & $+0.7$ & $-28 $ & $ 50$ & $-5\pm 7$\\
\ion{O}{vi} $\lambda 22.02$ & $+0.4$ & $-28 $ & $ 10$ & $-6\pm 5$\\
\ion{C}{v}  $\lambda 34.97$ & $+0.1$ & $-110$ & $100$ & $+8\pm 13$\\
\ion{O}{v}  $\lambda 22.37$ & $ 0.0$ & $-138$ & $ 40$ &$-11\pm 12$\\
\ion{N}{v}  $\lambda 29.41$ & $-0.1$ & $-166$ & $ 10$ &$-10\pm 15$\\
average &  & & & $-5.3\pm 3.6$ \\
\hline
\end{tabular}
\smallskip
\begin{list}{}{}
\item[$^{\mathrm{a}}$] ionisation parameter in $10^{-9}$~W\,m
\item[$^{\mathrm{b}}$] adopted velocity based on UV lines
\item[$^{\mathrm{c}}$] adopted systematic uncertainty in velocity based on 
the differences in centroids for the UV lines
\item[$^{\mathrm{d}}$] statistical uncertainty observed line centroid,
uncertainty in laboratory wavelength (Table~\ref{tab:lines}) and the adopted
systematic uncertainty in velocity have been added in quadrature 
\end{list}
\end{table}

The strong variation of average outflow velocity as a function of $\xi$ that is
apparent in Table~\ref{tab:uvoutflow}, in particular between \ion{O}{vi} and
\ion{N}{v}, shows that caution must be taken. Therefore, we consider here only
X-ray lines from ions which overlap in ionisation parameter with the ions in
Table~\ref{tab:uvoutflow}. We list those lines in Table~\ref{tab:xoutflow}.  For
\ion{C}{v} and \ion{O}{v} we have interpolated the velocity between the
\ion{O}{vi} and \ion{N}{v} velocity according to their $\log\xi$ value. The
weighted average shift $\Delta\lambda \equiv \lambda_{\mathrm{obs}} -
\lambda_{\mathrm{pred}}$ is $-5.3\pm 3.6$~m\AA.

\subsubsection{Lines from foreground neutral gas
absorption\label{sect:ismneutral}}

Our spectrum (see Table~\ref{tab:lines}) contains lines from the neutral as well
as the hot phase of the ISM. Here we discuss the neutral phase. The spectrum
shows two strong lines from this phase, the \ion{O}{i} and \ion{N}{i} is--2p
lines. Other lines are too weak to be used for wavelength calibration purposes.

The sight line to Mrk~509 is dominated by neutral material at LSR velocities of
about $+10$~km\,s$^{-1}$ and $+60$~km\,s$^{-1}$, as seen for instance in the
\ion{Na}{i} D$_1$ and D$_2$ lines, and the \ion{Ca}{ii} H and K lines
\citep{york1982}. Detailed 21~cm observations \citep{mcgee1986} show four
components: two almost equal components at $-6$ and $+7$~km\,s$^{-1}$, a weaker
component (6\% of the total \ion{H}{i} column density) at $+59$~km\,s$^{-1}$ and
the weakest component at $+93$~km\,s$^{-1}$ containing 0.7\% of the total column
density. 

The strongest X-ray absorption line from the neutral phase is the \ion{O}{i}
1s--2p transition. The rest-frame wavelength of this line has been determined as
$23.5113\pm 0.0018$~\AA\ \citep{kaastra2010}. Based on the velocity components
of \citet{mcgee1986}, we have estimated the average velocity for this line as
well as the UV line at 1302~\AA, assuming an oxygen abundance of $5.75\times
10^{-4}$ \citep[proto-Solar value: ][]{lodders2003}, adopting that $\sim 50$\%
of all neutral oxygen is in its atomic form. We predict for the
$\lambda$23.5~\AA\ line an LSR velocity of $+24$~km\,s$^{-1}$, and for the
$\lambda$1302~\AA\ line a velocity of $+40$~km\,s$^{-1}$. For a two times higher
column density, these lines shift by no more than $+6$ and $+1$~km\,s$^{-1}$,
respectively. For comparison, the measured centroid of the $\lambda$1302~\AA\
line \citep[Fig.~6]{collins2004} is $+40$~km\,s$^{-1}$, in close agreement with
the prediction. A systematic uncertainty of $\pm 10$~km\,s$^{-1}$ on the
predicted $\lambda$23.5~\AA\ line seems therefore to be justified. Taking this
into account, we expect the \ion{O}{i} line at $23.515\pm 0.002$~\AA.

\begin{table}[!ht]
\caption[]{Wavelengths (\AA) of the \ion{O}{iv} 1s--2p transitions; all
transitions are from the ground state to 1s2s$^2$2p$^2$.}
\smallskip
\label{tab:o4lines}
\begin{tabular}{ccccl}
\hline
$^2$D$_{3/2}$ & $^2$P$_{1/2}$ & $^2$P$_{3/2}$ & $^2$S$_{1/2}$ & reference \\
\hline
22.777 & 22.729 & 22.727 & 22.571 & HULLAC$^{\mathrm{a}}$ \\
22.755 & 22.739 & 22.736 & 22.573 & Cowan code$^{\mathrm{b}}$ \\
22.77  & 22.74  & 22.74  & 22.66  & AS2 (Autostructure)$^{\mathrm{c}}$ \\
22.78  & 22.75  & 22.75  & 22.65  & HF1 (Cowan code)$^{\mathrm{c}}$ \\
22.73  & \multicolumn{2}{c}{22.67} & 22.46 & R-matrix$^{\mathrm{d}}$ \\
  --   & \multicolumn{2}{c}{22.741$\pm$0.004} & -- & experiment$^{\mathrm{e}}$\\
\hline
22.777 & 22.741 & 22.739 & 22.571 & Adopted value \\
 0.010 &  0.004 &  0.004 &  0.025 & Adopted uncertainty \\
\hline
\end{tabular}
\smallskip
\begin{list}{}{}
\item[$^{\mathrm{a}}$] E. Behar 2001, private communication
\item[$^{\mathrm{b}}$] A.J.J. Raassen 2001, private communication
\item[$^{\mathrm{c}}$] \citet{garcia2005}
\item[$^{\mathrm{d}}$] \citet{pradhan2003}
\item[$^{\mathrm{e}}$] \citet{gu2005}
\end{list}
\end{table}

Unfortunately, the Galactic \ion{O}{i} $\lambda$23.5~\AA\ line is blended  by
absorption from lines of two other ions in the outflow of Mrk~509, \ion{O}{iv}
and \ion{Ca}{xv}. We discuss those lines below.

\paragraph{Contamination by \ion{O}{iv}:}

The \ion{O}{iv} 1s--2p\,$^2$P transition, which has a laboratory wavelength of
$22.741\pm 0.004$~\AA\ \citep{gu2005}, contaminates the Galactic \ion{O}{i}
line. In fact, the situation is more complex, as the main \ion{O}{iv} 1s--2p
multiplet has four transitions. In Table~\ref{tab:o4lines} we summarise the
theoretical calculations and experimental measurements of the wavelengths of
these lines. The adopted values are based on the single experimental measurement
and the HULLAC calculations, while the adopted uncertainties are based on the
scatter in the wavelength differences. 

In the present case, both the \ion{O}{iv} 1s--2p\,$^2$D$_{3/2}$ and
1s--2p\,$^2$P$_{1/2}$ and $^2$P$_{3/2}$ transitions of the outflow blend with
the Galactic \ion{O}{i} line, with relative contributions of 37, 42 and 21\%,
respectively. The precise contamination is hard to determine, as these are the
strongest \ion{O}{iv} transitions and the other \ion{O}{iv} transitions are too
weak to determine the column density. Based on the absorption measure
distribution of other ions with similar ionisation parameter, we estimate that
the total equivalent width of these three \ion{O}{iv} lines is $3.6\pm
2.0$~m\AA\ \citep{detmers2011}. The same modelling predicts that the \ion{O}{iv}
concentration peaks at $\log\xi = -0.6$, at almost the same ionisation parameter
as for \ion{C}{iv}. The optical depths of all X-ray lines of \ion{O}{iv} is less
than 0.1, hence we use the column density weighted average line centroid of
\ion{C}{iv} ($-197$~km\,s$^{-1}$) for \ion{O}{iv}. With this, we predict a
wavelength for the combined $^2$D and $^2$P transitions of \ion{O}{iv} of
$23.523\pm 0.004$~\AA.

\paragraph{Contamination by \ion{Ca}{xv}:}

The other contaminant to the Galactic \ion{O}{i} 1s--2p line is the
\ion{Ca}{xv} 2p--3d ($^3$P$_0$ -- $^3$D$_1$) $\lambda 22.730$~\AA\ transition of
the outflow. This is the strongest X-ray absorption line for this ion. Based on
the measured column density of \ion{Ca}{xiv} in the outflow, and assuming that
the \ion{Ca}{xv} column density is similar, we expect an equivalent width of
$4.9\pm 2.3$~m\AA\ for this line, about 10~\% of the strength of the \ion{O}{i}
blend.

The rest-wavelength of this transition is somewhat uncertain, however. 
\citet{kelly1987} gives a value of 22.725~\AA, without errors, and refers to
\citet{bromage1977} who state that their theoretical wavelength is within
5~m\AA\ from the observed value of 22.730~\AA, and that the line is blended.
Although not explicitly mentioned, their experimental value comes from
\citet{kelly1973} (cited in the Bromage \& Fawcett paper). \citet{kelly1973}
gives a value of 22.73~\AA, but states it is for the $^3$P$_2$ -- $^3$D$_3$
transition, which is a misclassification according to \citet{fawcett1975}. They
list as their source \citet{fawcett1971}. The latter publication gives the value
of 22.73, but no accuracy. In \citet{fawcett1970} the typical accuracy of the
plasma measurements is limited by Doppler motions of about $10^{-4}c$. If we
assume that accuracy also holds for the data in \citet{fawcett1971}, the
estimated accuracy should be about 2~m\AA. Given the rounding of
\citet{fawcett1970}, an r.m.s. uncertainty of 3~m\AA\ is more appropriate.
However, there is some level of blending with other multiplets. Blending is
probably due to other 2p--3d transitions in this ion, most likely transitions
between the $^3$P--$^3$D, $^3$P--$^3$P and $^1$D--$^1$F multiplets
\citep{aggarwal2003}. A comparison of the theoretical wavelengths of
\citet{aggarwal2003} with the experimental values of \citet{kelly1987} for all
lines of the $^3$P -- $^3$D multiplet shows, apart from an offset of 145~m\AA, a
scatter of 6~m\AA.

Given all this, we adopt here the original measurement of \citet{fawcett1971},
but with a systematic uncertainty of 5~m\AA. As \ion{Ca}{xv} has a relatively
large ionisation parameter ($\log\xi = 2.5$), we expect that like other highly
ionised species in Mrk~509 it is dominated by the high-velocity outflow
components, that have a velocity of about $-340$~km\,s$^{-1}$ (see
Table~\ref{tab:uvoutflow}). Putting a conservative  $200$~km\,s$^{-1}$
uncertainty on this (allowing a centroid up to halfway between the two main
absorption troughs), we may expect the line at $23.488\pm 0.017$~\AA.

\paragraph{Centroid of the \ion{O}{i} blend:}

The total equivalent width of the \ion{O}{i} blend is $29.1\pm 2.8$~m\AA\ (see
Table~\ref{tab:lines}). The velocity dispersion of the contaminating \ion{O}{iv}
and \ion{Ca}{xv} lines is an order of magnitude larger than the small velocity
dispersion of the cold Galactic gas, and both contaminants have optical depth
smaller than unity. The total blend is thus the combination of broad and shallow
contaminants with deep and narrow \ion{O}{i} components. Thus the contribution
due to \ion{O}{i} itself  is obtained by simply subtracting the contributions
from \ion{O}{iv} and  \ion{Ca}{xv} from the total measured equivalent width of
the blend. This yields an \ion{O}{i} equivalent width of $20.6\pm 4.2$~m\AA.
Taking into account the contribution of the contaminants, we expect the centroid
of the blend to be at $23.5115\pm 0.0032$~\AA. This corresponds to an offset
$\Delta\lambda = 9.9\pm 4.5$~m\AA.

\paragraph{\ion{N}{i}:}

The other strong X-ray absorption line from the neutral phase of the ISM is
the \ion{N}{i} 1s--2p line. We adopt the same LSR velocity as for the unblended
\ion{O}{i} line, $+24$~km\,s$^{-1}$. This is consistent with the measured line
profile of the \ion{N}{i} $\lambda 1199.55$~\AA\ line, which shows two troughs
at about 0 and $+50$~km\,s$^{-1}$ \citep{collins2004}. The shift derived from
the \ion{N}{i} line is therefore $+12.2\pm 8.9$~m\AA. 

We note, however, that for \ion{N}{i} we can only use RGS2, as RGS1 shows some
instrumental residuals near this line. The nominal wavelength for RGS1 is
26~m\AA\ higher than for RGS2. It is therefore worthwhile to do a sanity check.
Therefore we have determined the same numbers from archival data of Sco X-1.
That source was observed several times with significant offsets in the
dispersion direction, so that the spectral lines fall on different parts of the
CCDs or even on different CCD chips. We measure wavelengths for \ion{O}{i} and
\ion{N}{i} in Sco X-1 of $23.5123\pm 0.0015$ and $31.2897\pm 0.0030$~\AA,
respectively. Sco X-1 has a Heliocentric velocity of $-114$~km\,s$^{-1}$
\citep{steeghs2002}. Given its Galactic longitude of 0\fdg7, we expect the local
foreground gas to have negligible velocity. This is confirmed by an archival
HST/GHRS spectrum, taken from the HST/MAST archive, and showing a velocity
consistent with zero, to within $5$~km\,s$^{-1}$ for the \ion{N}{i} $\lambda
1199.55$~\AA\ line from the ISM. This then leads to consistent observed offsets
for the 1s--2p transitions of \ion{O}{i} and \ion{N}{i} in Sco X-1 of $1.0\pm
2.3$~m\AA\ and $4.0\pm 3.0$~m\AA, respectively.

Combining both the \ion{O}{i} and the \ion{N}{i} line for Mrk~509, we obtain a
shift of $+10.3\pm 4.0$~m\AA.

\subsubsection{Lines from foreground hot gas absorption}

\begin{table}[!ht]
\caption[]{Decomposition of the \ion{O}{vi} profile of \citet{sembach2003}
into Gaussian components.}
\smallskip
\label{tab:o6}
\begin{tabular}{crrr}
\hline
nr & v$_{\mathrm{LSR}}$ & $\sigma$          & column \\
   & (km\,s$^{-1}$)     & (km\,s$^{-1}$) & ($10^{16}$~m$^{-2}$) \\
\hline
1 & $-$245 & 59 & 214 \\
2 & $-$126 & 16 &  23 \\
3 & $-$64  & 13 &  19 \\
4 & $+$26  & 44 & 440 \\
5 & $+$152 & 24 &  26 \\
\hline
\end{tabular}
\smallskip
\end{table}

Our spectrum also contains significant lines from hot, ionised gas along the
line of sight, in particular from \ion{O}{vi}, \ion{O}{vii}, \ion{O}{viii}, and
\ion{C}{vi}. \citet{sembach2003} give a plot of the \ion{O}{vi} line profile.
This line shows clear high-velocity components at $-247$ and
$-143$~km\,s$^{-1}$. Unfortunately, no column density of the main Galactic
component is given by \citet{sembach2003}. Therefore we have fitted their
profile to the sum of 5 Gaussians (Table~\ref{tab:o6}). The column-density
weighted centroid for the full line is at $-56$~km\,s$^{-1}$ in the LSR frame.
We estimate an uncertainty of about 8~km\,s$^{-1}$ on this number.

Note that a part of the high-velocity trough in the blue is contaminated by a
molecular hydrogen line at 1031.19~\AA, which falls at $v_{\rm{LSR}} =
-206$~km\,s$^{-1}$. This biases component 1 of our fit for the hot foreground
galactic component by about 10~km\,s$^{-1}$. This component in \ion{C}{iv},
\ion{Si}{iv}, and \ion{N}{v} in the COS spectra is at a velocity of $v_{\rm
{LSR}}=-245$~km\,s$^{-1}$ \citep{kriss2011}, and therefore we adopt that value
here.

\begin{table}[!ht]
\caption[]{Line shifts $\Delta\lambda \equiv \lambda_{\mathrm{obs}} -
\lambda_{\mathrm{pred}}$ in m\AA\ for the X-ray absorption lines from the hot
interstellar medium.}
\smallskip
\label{tab:uvhot}
\begin{tabular}{lcccc}
\hline
transition & $\Delta\lambda$$^{\mathrm{a}}$ & $\Delta\lambda$$^{\mathrm{b}}$ &
error$^{\mathrm{c}}$ & error$^{\mathrm{d}}$\\
   & (m\AA) & (m\AA)    & (m\AA) & (m\AA) \\
\hline
\ion{Fe}{xvii} 2p--3d &  9    &  5   &  8 & 13 \\
\ion{Ne}{x}    1s--2p & $-$8  &$-$11 & 10 & 13 \\
\ion{Ne}{ix}   1s--2p &  9    &  6   &  7 & 11 \\
\ion{O}{vii}   1s--3p & $-$3  &$-$8  & 10 & 16 \\
\ion{O}{viii}  1s--2p &  0    & $-$5 &  5 & 14 \\
\ion{O}{vii}   1s--2p & $-2$  & $-$8 &  5 & 15 \\
\ion{O}{vi}    1s--2p &  8    &  2   & 13 & 13 \\
\ion{C}{vi}    1s--2p & 11    &  1   & 13 & 26 \\
\hline
\end{tabular}
\smallskip
\begin{list}{}{}
\item[$^{\mathrm{a}}$] Using $v_{\mathrm LSR}=-56$~km\,s$^{-1}$
\item[$^{\mathrm{b}}$] Using $v_{\mathrm LSR}=+26$~km\,s$^{-1}$
\item[$^{\mathrm{c}}$] Only statistical error
\item[$^{\mathrm{d}}$] Including uncertainty of $\pm 200$~km\,s$^{-1}$ except
for \ion{O}{vi}
\end{list}
\end{table}

While this velocity can be readily used for predicting the observed wavelength
of the \ion{O}{vi} 1s--2p line, the uncertainty for other X-ray lines is larger,
because the temperatures or ionisation states of the dominant components 1 and 4
may be different. Moreover, \ion{O}{vi} is the most highly ionised ion available
in the UV band, while most of our X-ray lines from the ISM have a higher degree
of ionisation. Therefore, without detailed modelling of all these components, we
conservatively assume here for all ions other than \ion{O}{vi} an error of
200~km\,s$^{-1}$ on the average outflow, corresponding to the extreme cases that
the bulk of the column density is in either component 1 or 5.

The Galactic \ion{O}{vii} 1s--2p line at 21.6~\AA\ suffers from some blending by
the \ion{N}{vii} 1s--3p line from the outflow. That last line, with a predicted
equivalent width of 4.5~m\AA, has a relatively high ionisation parameter,
$\log\xi = 1.4$. Therefore we do not know exactly what its predicted velocity
should be, and we conservatively assume a value of $-100\pm 200$~km\,s$^{-1}$,
spanning the full range of outflow velocities. With that, the predicted
wavelength of the \ion{N}{vii} outflow line is $21.624\pm 0.014$~\AA, while
the predicted wavelength of the \ion{O}{vii} line is $21.5993\pm 0.0006$~\AA.  
The predicted wavelength of the blend then is $21.606\pm 0.004$~\AA.

In Table~\ref{tab:uvhot} we list the lines that we use and the derived line
shifts $\Delta\lambda$. The weighted average is $+1.7\pm 2.6$~m\AA\ when we only
include statistical errors, and $+2.8\pm 4.9$ ~m\AA\ if we include the
systematic uncertainties.

We make here a final remark. Combining all wavelength scale indicators including
the one above (Sect.~\ref{sect:wavsummary}), we obtain an average shift of
$-1.6$~m\AA. Using this number, the most significant lines of
Table~\ref{tab:uvhot} (\ion{O}{vii} and \ion{O}{viii} 1s--2p) show clearly a
velocity much closer to the dominant velocity component 4 ($+26$~km\,s$^{-1}$,
Table~\ref{tab:o6}) than to the column-density weighted velocity of the
\ion{O}{vi} line at $-56$~km\,s$^{-1}$. This last number is significantly
affected by the high-velocity clouds, that apparently are less important for the
more highly ionised ions.  Therefore, in Table~\ref{tab:uvhot} we also list in
column 3 the wavelength shifts under the assumption that all lines except
\ion{O}{vi} are dominated by component 4. With that assumption, and taking now
only the statistical errors into account, the weighted average shift is $-1.7\pm
2.8$~m\AA.

\subsubsection{Comparison with a sample of RGS spectra of stars}

Preliminary analysis (R. Gonz\'alez-Riestra, private communication) of a large
sample of RGS spectra of stellar coronae indicate that there is a correlation
between the wavelength offsets and the Solar Aspect Angle (SAA) of the satellite
(correlation coefficient about 0.6). At a mean SAA of 90\degr, the average
$\Delta\lambda$ for RGS1 is 2.3~m\AA, while for RGS2 it is 7.3~m\AA. The scatter
for an individual observation is 6 and 7~m\AA, respectively. The difference
between RGS1 and RGS2 is fairly consistent with the $6.9\pm 0.7$~m\AA\ that we
adopted for the present work. As our observation campaign on Mrk~509 spans the
full visibility window, the average SAA angle is 90\degr; it varied between
108\fdg5 for the first observation, to 73\fdg3 for the last
observation. The wavelengths in our spectrum (averages of the original RGS1
wavelengths and the RGS2 wavelengths minus 6.9~m\AA) should then be off by
$+1.4\pm 6.5$~m\AA.

The above estimate can be refined further. It appears that there is also a
correlation of the wavelength offset with the temperature parameters T2007 and
T2031 of the grating boxes on RGS1 and RGS2, respectively. However, these
temperatures do not correlate well with the SAA. Hence, following R.
Gonz\'alez-Riestra et al. (private communication) and using their data on 38
spectra of stellar coronae, we obtain the following correlations:
\begin{eqnarray}
\label{eqn:double}
\mathrm{RGS1}:\  \Delta\lambda = 
      4.92 &-& 0.49 (\mathrm{SAA}-90\degr) \nonumber \\
           &-& 12.8 (\mathrm{T2007} - 20.7), \\
\mathrm{RGS2}:\  \Delta\lambda = 
      8.32 &-& 0.55 (\mathrm{SAA}-90\degr) \nonumber \\
           &-& 11.8 (\mathrm{T2031} - 20.7),
\label{eqn:double2}
\end{eqnarray}
where SAA is expressed in degrees, and the temperatures in {\degr}C.
The scatter around this relation is 4.1 and 4.8~m\AA\ for RGS1 and RGS2,
respectively. This scatter is significantly smaller than the scatter using only
the correlation with SAA mentioned above.

\begin{table}[!ht]
\caption[]{Average line shifts for the six strongest lines,
relative to the total spectrum.}
\smallskip
\label{tab:dwi}
\begin{tabular}{cccccc}
\hline
\smallskip
Obs & $\lambda - \overline{\lambda}$$^{\mathrm{a}}$ & SAA & T2007 & T2031 & 
   $\lambda - \overline{\lambda}$$^{\mathrm{b}}$ \\
 & (m\AA) & (\degr) & ($^{\mathrm{o}}$C) & ($^{\mathrm{o}}$C) & (m\AA)\\
\hline
 1 & $ -6.2\pm 3.0$ & 108.5 &  21.38 &  21.48 &  -10.60 \\
 2 & $-12.2\pm 2.9$ & 104.1 &  21.26 &  21.38 &   -6.97 \\
 3 & $ +3.9\pm 1.9$ & 100.7 &  21.32 &  21.42 &   -5.80 \\
 4 & $ +2.4\pm 2.2$ &  94.7 &  21.24 &  21.33 &   -1.68 \\
 5 & $ -3.6\pm 2.1$ &  90.9 &  21.43 &  21.51 &   -1.98 \\
 6 & $ -3.0\pm 2.1$ &  86.7 &  21.51 &  21.54 &   -0.48 \\
 7 & $  0.0\pm 2.3$ &  82.7 &  21.28 &  21.40 &    3.90 \\
 8 & $ +7.3\pm 2.3$ &  78.7 &  21.32 &  21.43 &    5.52 \\
 9 & $ +7.6\pm 2.1$ &  75.0 &  21.24 &  21.35 &    8.42 \\
10 & $ +0.3\pm 2.3$ &  72.7 &  21.24 &  21.34 &    9.68 \\
\hline
all$^{\mathrm{c}}$ &   $\equiv 0$   &  89.5 &  21.32 &  21.42 & $\equiv 0$ \\
\hline
\end{tabular}
\smallskip
\begin{list}{}{}
\item[$^{\mathrm{a}}$] Observed wavelength difference for individual
observations.
\item[$^{\mathrm{b}}$] Predicted wavelength difference for individual
observations using (\ref{eqn:double})--(\ref{eqn:double2}); results for RGS1 and RGS2 have been
weighted averaged (not all lines present in each RGS).
\item[$^{\mathrm{c}}$] Combined spectrum.
\end{list}
\end{table}

\begin{figure}
\resizebox{\hsize}{!}{\includegraphics[angle=-90]{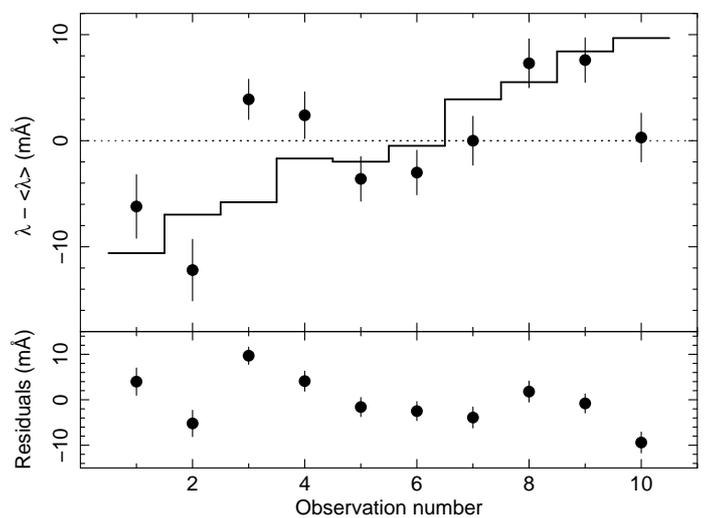}}
\caption{Top panel: wavelength difference for the six strongest lines
in each observation relative to the lines in the average spectrum (data points)
and model based on Eq.~(\ref{eqn:double})--\ref{eqn:double2} (histogram). Bottom panel: residuals
of the data points relative to the model in the upper panel. }
\label{fig:dw}
\end{figure}

We apply this now to the 10 individual observations of Mrk~509. We have
determined the individual line centroids of the 6 strongest lines (\ion{C}{vi}
1s--2p, \ion{O}{viii} 1s--2p and 1s--3p, \ion{O}{vii} 1s--2p, \ion{Ne}{ix}
1s--2p all from the outflow, and \ion{O}{i} 1s--2p from the foreground). We then
take the difference with the wavelength in the combined observation, and
averaged these residuals. We show the results in Table~\ref{tab:dwi} and
Fig.~\ref{fig:dw}.

The model gives an improvement: the r.m.s. residuals (after correction for the 
mean statistical errors on the data points of 2.3~m\AA) are reduced from 5.4 to
4.6~m\AA. This is fully consistent with the remaining residuals in the set of
coronal spectra discussed above. It is also clear that the remaining residuals
are non-statistical, and although on average the model gives an improvement, in
some individual cases the opposite is true, for instance for observations 3, 7
and 10.

Keeping this in mind, we can use the model to predict the absolute wavelength
scale. We predict absolute offsets $\Delta\lambda$ of $-2.76$ and $+0.10$~m\AA\
for RGS1 and RGS2, respectively. In our extraction procedure, we have subtracted
6.9~m\AA\ from the RGS2 wavelengths; doing the same for the predicted RGS
offset, and averaging the offsets for RGS1 and RGS2, the model predicts a total
offset of $-4.8$~m\AA. We only need to apply a small correction for the velocity
of the Earth. It can be assumed that the sample of stars has, on average, zero
redshift due to the motion of Earth, because usually each source can be observed
twice a year with opposite signs of the Doppler shift. Our data of Mrk~509 were
taken with a rather extreme Earth velocity of $-29.3$~km\,s$^{-1}$. At a typical
wavelength of 20~\AA, this corresponds to a correction factor of 2.0~m\AA. We
finally obtain a predicted offset of $-2.8\pm 4.5$~m\AA\, where the error is the
average systematic residual found for the sample of coronae mentioned before. 

\subsubsection{Comparison with Chandra LETGS spectra}

\begin{table}[!ht]
\caption[]{Comparison of measured wavelengths from the RGS and LETGS.}
\smallskip
\label{tab:letgs}
\begin{tabular}{cccc}
\hline
transition & RGS & LETGS & RGS $-$ LETGS \\
 & (\AA) & (\AA) & (m\AA) \\
\hline
\ion{O}{viii}  1s--3p & $16.5460\pm 0.0023$ & $16.552\pm 0.004$ & $-6\pm 5$ \\
\ion{O}{vii}   1s--3p & $19.2649\pm 0.0029$ & $19.269\pm 0.007$ & $-4\pm 8$ \\  
\ion{O}{viii}  1s--2p & $19.6146\pm 0.0012$ & $19.619\pm 0.004$ & $-4\pm 4$ \\  
\ion{O}{vii}   1s--2p & $22.3354\pm 0.0018$ & $22.346\pm 0.004$ & $-11\pm 4$ \\ 
\hline
\end{tabular}
\smallskip
\end{table}

We have also measured the wavelengths of the 1s--2p and 1s--3p lines of
\ion{O}{vii} and \ion{O}{viii} in the Chandra LETGS spectrum of Mrk~509 taken 20
days after our last XMM-Newton observation. LETGS has the advantage above RGS
that it measures both positive and negative spectral orders, hence allow to
determine the zero-point of the wavelength scale more accurately.
Table~\ref{tab:letgs} lists the wavelengths measured by both instruments. The
average wavelength offset of RGS relative to LETGS is $-6.9\pm 2.4$~m\AA. The
above number only includes the statistical uncertainties. We have extracted an
LETGS spectrum of Capella, taken around the same time as the Mrk~509
observation. The ten strongest lines in the spectrum of Capella between
8--34~\AA\ show a scatter with r.m.s. amplitude of 6~m\AA\ compared to the
laboratory wavelengths, but no systematic offset larger than 2~m\AA. We
therefore add a systematic uncertainty of $6/\sqrt{4}=3$~m\AA\ in quadrature to
the statistical uncertainty of the average $\Delta\lambda$ for the four lines in
Mrk~509 that we use here.

We only need to make a small correction of $-0.4$~m\AA\ corresponding to the
higher speed of Earth away from Mrk~509 ($-29.3$~km\,s$^{-1}$) during the
XMM-Newton observation as compared to the Chandra observation
($-23.3$~km\,s$^{-1}$). This all leads to an offset of $-7.3\pm 3.8$~m\AA.

\subsubsection{Summary of results on the wavelength scale
\label{sect:wavsummary}}

\begin{table}[!ht]
\caption[]{Overview of wavelength scale indicators.}
\smallskip
\label{tab:woverview}
\begin{tabular}{l@{\hspace{-0.5truecm}}r}
\hline
Indicator & $\Delta\lambda \equiv \lambda_{\mathrm{obs}} -
\lambda_{\mathrm{pred}}$ \\
 & (m\AA) \\
\hline
1. Lines from the outflow of Mrk~509                       &  $-5.3\pm 3.6$ \\
2. Lines from foreground neutral gas$^{\mathrm{a}}$        &  $+10.3\pm 4.0$ \\
2. Lines from foreground neutral gas$^{\mathrm{b}}$        &  $+6.5\pm 3.4$ \\
3. Lines from foreground hot gas absorption$^{\mathrm{c}}$ &  $+2.8\pm 4.9$ \\
3. Lines from foreground hot gas absorption$^{\mathrm{d}}$ &  $-1.9\pm 4.9$ \\
4. Comparison with a sample of RGS spectra of stars        &  $-2.8\pm 4.5$ \\
5. Comparison with Chandra LETGS spectra                   &  $-7.3\pm 3.8$ \\
\hline
All indicators combined$^{\mathrm{a,c}}$                &  $-1.0\pm 1.8$ \\
All indicators combined$^{\mathrm{b,c}}$                &  $-1.2\pm 1.8$ \\
All indicators combined$^{\mathrm{a,d}}$                &  $-1.6\pm 1.8$ \\
All indicators combined$^{\mathrm{b,d}}$                &  $-1.8\pm 1.8$ \\
\hline
\end{tabular}
\smallskip
\begin{list}{}{}
\item[$^{\mathrm{a}}$] As described in Sect.~\ref{sect:ismneutral}
\item[$^{\mathrm{b}}$] Ignoring the blending by \ion{Ca}{xv} of the \ion{O}{i}
line
\item[$^{\mathrm{c}}$] Using the \ion{O}{vi} centroid velocity
\item[$^{\mathrm{d}}$] Using the velocity of dominant component 4 in \ion{O}{vi}
\end{list}
\end{table}

We now summarise our findings from the previous sections
(Table~\ref{tab:woverview}). All estimates agree within the error bars, except
for the lines from the neutral foreground gas, which is off by about 3$\sigma$.
This number is dominated by the rather complex \ion{O}{i} blend. Possibly there
are larger systematic uncertainties on this number than we assumed. The major
factor here is the blending with \ion{Ca}{xv}. If we assume that the equivalent
width of the contaminating \ion{Ca}{xv} line is zero, the predicted blend
centroid decreases by 4.4~m\AA, towards better agreement with the other
indicators. The \ion{O}{iv} blend has a much smaller impact, as its centroid is
already close to the centroid of the Galactic \ion{O}{i} line.

We note that a smaller value of the predicted equivalent width of the
\ion{Ca}{xv} line may be caused by a somewhat lower Ca abundance; alternatively,
if the density is higher than about $10^{16}$~m$^{-3}$, the population of
\ion{Ca}{xv} in the ground state decreases significantly \citep{bhatia1993},
leading also to a reduced equivalent width of this absorption line.

All five methods used here have their pros and cons, and we have done the best
we can to estimate the uncertainties for each of the methods. The five methods
are for the major part statistically independent, and this allows us to use a
weighted average for the final wavelength correction. Ignoring the blending by
\ion{Ca}{xv} to the Galactic \ion{O}{i} line, and using $+26$~km\,s$^{-1}$ for
the outflow velocity of the hot Galactic gas, this wavelength shift is $-1.8\pm
1.8$~m\AA. Within their statistical uncertainties, all individual indicators are
consistent with this value, except for the lines from foreground neutral gas,
which are off by $8.3\pm 3.8$~m\AA. We do not have an explanation for that
deviation.

In our analysis, we apply the wavelength corrections by shifting the wavelength
grids in our program {\sl rgs\_fmat}.

\section{AGN outflow dynamics\label{sect:agn}}

\begin{figure}
\resizebox{\hsize}{!}{\includegraphics[angle=-90]{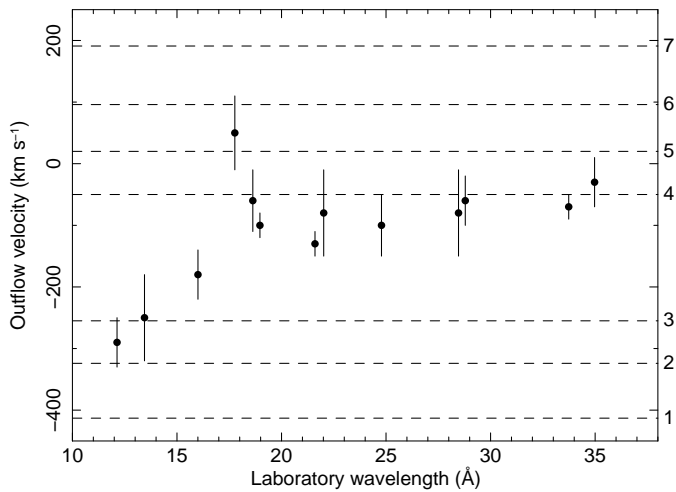}}
\caption{Average outflow velocity as measured through individual lines
with statistical errors smaller than 100~km\,s$^{-1}$. The dashed lines show
the 7 velocity components as seen in archival FUSE data \citep{kriss2000}.}
\label{fig:v_outflow}
\end{figure}

To show how the optimised wavelength calibration improves the scientific
results, we give here a brief analysis of the velocity structure of the AGN
outflow as measured in X-rays through two Rydberg series of oxygen ions. A full
analysis of the spectrum is given by \cite{detmers2011}.

Using the measured outflow velocities for individual lines, we can immediately
derive some interesting astrophysical conclusions (Fig.~\ref{fig:v_outflow}).

Most of the lines are in between the high-velocity components 1--3 and the lower
velocity components 4--7 as found in the \ion{O}{vi} lines \citep{kriss2000}.
This shows that there are multiple velocity components present in the X-ray
absorption lines. At longer wavelengths these lines are closer to the low
velocity components, showing that there is more column density in those
components. However, for the shortest wavelength lines, in particular the 1s--2p
transitions of \ion{Ne}{ix} and \ion{Ne}{x}, the centroid is in the range of the
high-velocity components 1--3. As the shorter wavelength lines originate from
ions with a higher ionisation parameter, there is a correlation between the
ionisation parameter and the outflow velocity.

The relative deviation of the 17.8~\AA\ line (\ion{O}{vii} 1s--4p) with respect
to the other lines, including the 1s--2p and 1s--3p lines of the same ion, can
be easily explained: according to preliminary modelling, the optical depth at
line centre for the 1s--2p, 1s--3p and 1s--4p lines of \ion{O}{vii} is 8, 1.4
and 0.5, respectively. Thus, if the \ion{O}{vii} column density of the low
velocity component is larger than that of the high-velocity component, the
strongest low-velocity line components (like 1s--2p, 1s--3p) will saturate while
the high-velocity components of these transitions will not saturate, hence
effectively these strong lines will be blueshifted relative to the weaker 1s--4p
line, consistent with what is observed.

We have elaborated on the line diagnostics for two of the most important ions in
the outflow of Mrk~509 to see if we can detect both velocity components:
\ion{O}{viii} and \ion{O}{vii}.

\begin{table}[!ht]
\caption[]{Parameters of the velocity components for \ion{O}{viii} in the
outflow of Mrk~509. Numbers in parenthesis were kept fixed.}
\smallskip
\label{tab:o8fit}
\begin{tabular}{lccc}
\hline
Parameter                          & Model 1 & \multicolumn{2}{c}{Model 2} \\
                                   &              & comp. 1--3  & comp. 4--7\\
\hline
$v$ (km\,s$^{-1}$)                 & $-115\pm 17$ & ($-329$) & ($+46$) \\
$\sigma$ (km\,s$^{-1}$)            & $113\pm 7$   & $45\pm 8$& $71\pm 15$ \\
$N_{\rm ion}$ ($10^{20}$~m$^{-2}$) & $10.7\pm 1.3$& $6.9\pm 2.6$ &$4.3\pm 1.5$\\
                      && \multicolumn{2}{c}{total: $ 11.2\pm 2.0$} \\
$\chi^2$                           & 4.21         &\multicolumn{2}{c}{3.01}\\
d.o.f.                             & 5            &\multicolumn{2}{c}{4}\\
\hline
\end{tabular}
\end{table}

\begin{figure}
\resizebox{\hsize}{!}{\includegraphics[angle=-90]{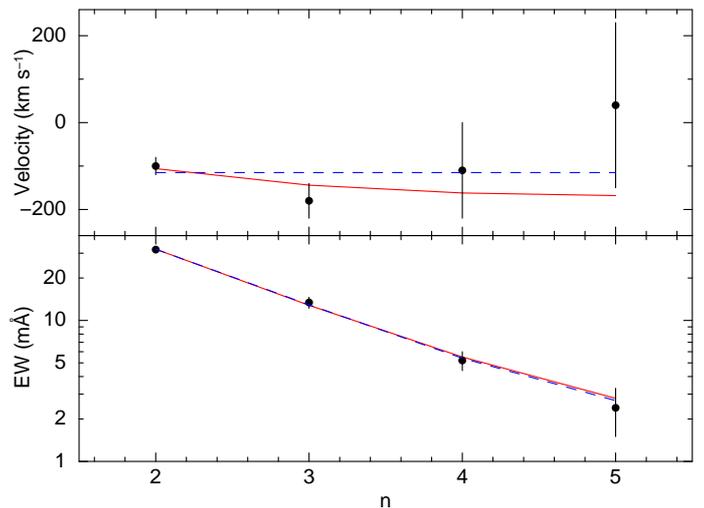}}
\caption{Equivalent width and line centroid for the 1s--2p to 1s--5p lines
of \ion{O}{viii}, plotted versus principal quantum number $n$ of the upper 
level. The predicted values for model 1 are shown as dashed lines,
those for model 2 as solid lines.}
\label{fig:o8mod}
\end{figure}

For \ion{O}{viii} we have measured velocities and equivalent widths for the
1s--2p to 1s--5p lines (Table~\ref{tab:lines}). We have tried different models.
The first model consists of a single Gaussian absorption line, with free
centroid, width and column density. Model 2 has two Gaussian components, with
the centroids frozen to the values given in Table~\ref{tab:uvoutflow} as derived
from the \ion{O}{vi} troughs, but with the widths and column densities free
parameters. The results are shown in Table~\ref{tab:o8fit} and
Fig.~\ref{fig:o8mod}. Both models give a good fit and cannot be distinguished on
statistical grounds. The total column densities for both models are the same
within their error bars, and also the sum of the width of the Gaussians for
Model 2 is consistent with the width of the single Gaussian for Model 1.
Apparently, as far as the determination of total line width and column density
is concerned, the detailed column density distribution as a function of velocity
is not very important. This can be understood as the strongest line with the
best statistics (1s--2p) has a large optical depth (12.7 and 4.9 for the two
components in model 2). The line core is thus black, the line shape rather
``rectangular'' and to lowest order approximation the width of the line is given
by the sum parts of the spectrum where the core is black.

For Model 2, the total width for each of the components agrees well with the
total width of components 1--3 and 4--7 of the \ion{O}{vi} line. Thus,
\ion{O}{viii} originates from multiple components, but the fastest outflow
components have a larger column density than the slower outflow components.

\begin{table}[!ht]
\caption[]{Parameters of the velocity components for \ion{O}{vii} in the
outflow of Mrk~509. Numbers in parenthesis were kept fixed. }
\smallskip
\label{tab:o7fit}
\begin{tabular}{lccc}
\hline
Parameter                          & Model 1 & \multicolumn{2}{c}{Model 2} \\
                                   &              & comp. 1--3  & comp. 4--7\\
\hline
$v$ (km\,s$^{-1}$)                 & $-102\pm 17$ & ($-329$) & ($+46$) \\
$\sigma$ (km\,s$^{-1}$)            & $ 96\pm 9$   & $80(>60)$& $51\pm 8$\\
$N_{\rm ion}$ ($10^{20}$~m$^{-2}$) & $5.9\pm 0.7$& $0.7(-0.1,+0.6)$ &$6.9\pm 1.6$\\
                      && \multicolumn{2}{c}{total: $  7.8\pm 1.4$} \\
$\chi^2$                           & 22.57        &\multicolumn{2}{c}{12.93}\\
d.o.f.                             & 5            &\multicolumn{2}{c}{4}\\
\hline
\end{tabular}
\end{table}

We have done a similar analysis for the 1s--2p to 1s--5p lines of \ion{O}{vii}
(Table~\ref{tab:o7fit}). The velocity broadening is consistent with what we
obtained for \ion{O}{viii}, although there is somewhat larger uncertainty. For 
\ion{O}{vii}, however, the low velocity components 4--7 have clearly the highest
column density, similar to what was found for \ion{O}{vi} \citep{kriss2000}. 

\section{Statistics\label{sect:statistics}}

Local continuum fits to RGS spectra in spectral bands that have few lines yield
in general $\chi^2$ values that are too low. This is an artifact of the SAS
procedures related to the rebinning of the data. Data have to be binned from the
detector pixel grid to the fixed wavelength grid that we use in our analysis.
However, the bin boundaries of both grids do not match. As a consequence of this
process, the natural Poissonian fluctuations on the spectrum as a function of
detector pixel are distributed over the wavelength bin coinciding most with the
detector pixel and its neighbours. In addition, there is a small smoothing
effect caused by pointing fluctuations of the satellite. Due to this tempering
of the Poissonian fluctuations, $\chi^2$ values will be lower for a ``perfect''
spectral fit.

We have quantified the effect by fitting a linear model $F_\lambda = a +
b\lambda$ to fluxed spectra in relatively line-poor spectral regions, in 1~\AA\
wide bins in the 7--10, 17--18 and 35--38~\AA\ ranges. The median reduced
$\chi^2$ is 0.72, with a 67~\% confidence range between 0.65 and 0.79. Over the
full 7--38~\AA\ range, we have at least 400 counts (and up to 3000 counts) in
each bin of $\sim$0.02~\AA\ width. Therefore, the Poissonian distribution is
well approximated by the Gaussian distribution, and the usage of chi-squared
statistics is well justified.

The rebinning process conserves the number of counts, hence the nominal error
bars (square root of the number of counts) are properly determined. The lower
reduced $\chi^2$ is caused by the smoothing effect on the data. For correct
inferences about the spectrum, such a bias in $\chi^2$ is not appropriate. As we
cannot change the observed flux values, we opt to multiply the nominal errors on
the fluxes by $\sqrt{0.72} = 0.85$, in order to get acceptable values for ideal
data and models. 

\begin{figure}
\resizebox{\hsize}{!}{\includegraphics[angle=-90]{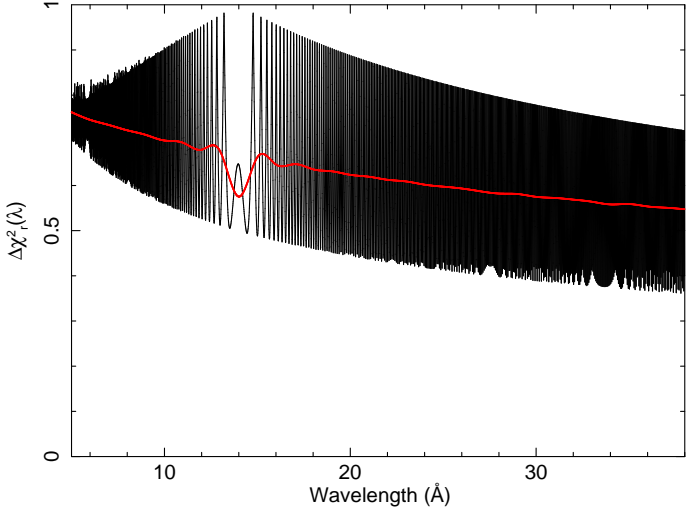}}
\caption{Relative contribution per wavelength bin to the reduced $\chi^2$.
Calculated and shown is the ratio ${\rm Var}[Y_i] / {\rm E}[Y_i]$ (see text).
The thick line is a low-resolution spline approximation showing 
the trends in the average value.}
\label{fig:chir}
\end{figure}

As a final check we have made a more robust estimate of the effect. For RGS1,
the wavelength channel number $c_\lambda$ as a function of dispersion angle
channel $c_\beta$ can be approximated well with
\begin{equation}
c_\lambda = 5.3708506 + 0.66942769 c_\beta + 1.12824091\times 10^{-4} c_\beta^2.
\label{eqn:disp}
\end{equation}
Locally, if the spectrum on the $\beta$-grid is given by a set of random
variables $X_j$ and the spectrum on the wavelength grid is given by
$Y_i$, then we have
\begin{equation}
Y_i = \sum\limits_{j}^{} f_{ij} X_j,
\end{equation}
for a small number of bins $j$ around the corresponding $\beta$-channel given by
(\ref{eqn:disp}). It is here important to note that while the $X_j$
variables are statistically independent, the $Y_i$ variables are in general {\sl
not} statistically independent. Typically, the $f_{ij}$ factors are proportional to the
overlap of wavelength bin $j$ with $\beta$-bin $i$. The clue is now that from
elementary statistical considerations we have for the expected value of $Y_i$
and the variance
\begin{equation}
{\rm E}[Y_i] = \sum\limits_{j}^{} f_{ij} {\rm E}[X_j],
\end{equation}
\begin{equation}
{\rm Var}[Y_i] = \sum\limits_{j}^{} f_{ij}^2 {\rm Var}[X_j].
\end{equation}
For instance, when the single $\beta$-bins $j$ would be exactly redistributed
over two adjacent $\lambda$-bins, then there are two $f$-values, each $=0.5$,
and we have ${\rm Var}[Y_i] / {\rm E}[Y_i] = 0.5 {\rm Var}[X_j] / {\rm E}[X_j]$,
where the factor 0.5 comes from $0.5^2+0.5^2$. We have done the calculation
exactly using the grids related through (\ref{eqn:disp}), and show our result in
Fig.~\ref{fig:chir}.

\section{Conclusion}

\subsection{Wavelength scale and spectral resolution}

By using five different methods, we have obtained a consistent wavelength scale
with an accuracy of about 1.8~m\AA, four times better than the nominal
wavelength accuracy of RGS. The only slightly deviant diagnostic remains the
neutral gas in our galaxy (\ion{O}{i} and \ion{N}{i}), which are off from the
mean by $8.8\pm 3.8$~m\AA\ (2.3$\sigma$). The systematics due to blending (for
\ion{O}{i}) and the lack of sufficiently reliable RGS data for \ion{N}{i} may be
slightly larger than we have estimated.

Our work shows that in cases like ours a dedicated effort to improve the
wavelength scale is rewarding. As we showed, at this level of accuracy, Doppler
shifts due to the orbit of Earth around the Sun become measurable. It also
allows us to use X-ray lines as tools to study the dynamics of the outflow and
the Galactic foreground at a level beyond the nominal resolution of the
instrument (see Sect.~\ref{sect:agn}).

\subsection{Effective area corrections}

The corrections that we have derived for the relative effective area of both RGS
detectors, as well as for the relative effective area of both spectral orders
are at the few percent level but should be taken into account for the high
statistical quality spectrum that we consider here. This is due to the fact that
there are data gaps for each individual RGS, but each time at a different
wavelength range. Without such a correction, there would be artificial jumps at
the boundaries of these regions of the order of a few percent. This leads any
spectral fitting package to attempt to compensate for this by enhancing or
decreasing astrophysical absorption lines or edges, in the hunt for the lowest
$\chi^2$ value.

A disadvantage of our method is of course that the ``true'' effective area and
hence absolute flux is known only to within a few percent, and the deviations
are wavelength-dependent. In our subsequent spectral fitting \citep{detmers2011}
we compensate for this by using a spline for the AGN continuum. This is clearly
justified, as one might worry about the intrinsic applicability of e.g. pure
power-law or blackbody components when the flux can be measured down to the
percent level accuracy. The ``true'' continua may contain all kinds of features
including e.g. relativistically broadened lines
\citep[e.g.][]{branduardi2001,sako2003} that can be discerned only in the
highest-accuracy data, such as the present one.

\subsection{Combination of spectra}

The reduction of the combined response matrices of our 40 spectra with 2 Gb of
memory into a single spectrum and response file of only 8 Mb, a reduction by
more than two orders of magnitude, allows us to make complex spectral fits. In
the most sophisticated models used by \citet{detmers2011} the number of free
parameters approaches 100, and without this efficient response matrix the
analysis would have been impossible.

\subsection{Final remarks}

Due to the high statistical quality of our data, the binning problems mentioned
in Sect.~\ref{sect:statistics} became apparent. In many RGS spectra that have
been obtained during the lifetime of XMM-Newton, the statistical quality of the
spectra is not good enough to detect this effect. Still, this could imply that
when users obtain a fit with reduced chi-squared of order unity, the fit is
formally not acceptable because an ideal fit would yield a smaller reduced
chi-squared. Clearly, here more work needs to be done to alleviate the problem,
but due to the discretisation of wavelength in the detector corresponding to the
natural size of the CCD pixels there is some natural limitation to what can be
done. Even in the best observations there is always some small scatter in the
satellite pointing, making some rebinning necessary.

Overall, however, the calibration of RGS is at a fairly advanced stage. Only for
exceptionally good spectra like the Mrk~509 spectrum that we discussed here the
sophisticated tools developed in this paper are needed, although also for weaker
sources our improvements can be beneficial for the analysis. Our tools are
publicly available as auxiliary programs ({\sl rgsfluxcombine}, {\sl rgs\_fmat})
in the SPEX distribution.

\begin{acknowledgements}

We thank Charo Gonz\'alez-Riestra (ESA) for providing the data on the RGS
spectra of stars in the context of the wavelength scale calibration. This work
is based on observations obtained with XMM-Newton, an ESA science mission with
instruments and contributions directly funded by ESA Member States and the USA
(NASA). SRON is supported financially by NWO, the Netherlands Organization for
Scientific Research. K. Steenbrugge acknowledges the support of Comit\'e Mixto
ESO - Gobierno de Chile. Ehud Behar was supported by a grant from the ISF. G.
Kriss gratefully acknowledges support from NASA/XMM-Newton Guest Investigator
grant NNX09AR01G. M. Mehdipour acknowledges the support of a PhD studentship
awarded by the UK Science \& Technology Facilities Council (STFC). P.-O.
Petrucci acknowledges financial support from CNES and the French GDR PCHE.
Ponti acknowledges support via an EU Marie Curie Intra-European Fellowship under
contract no. FP7-PEOPLE-2009-IEF-254279. G. Ponti and S. Bianchi acknowledge
financial support from contract ASI-INAF n. I/088/06/0.

\end{acknowledgements}

\bibliographystyle{aa}
\bibliography{rgsanal}

\end{document}